\documentclass[12pt]{iopart}

\usepackage{amsmath}
\usepackage{iopams}  
\usepackage{amsbsy}
\usepackage{amssymb}
\usepackage{graphicx}
\usepackage{subfig}
\usepackage[normalem]{ulem}
\usepackage[colorlinks=true, allcolors=blue]{hyperref}

\DeclareMathOperator*{\SumInt}{%
\mathchoice%
  {\ooalign{$\displaystyle\sum$\cr\hidewidth$\displaystyle\int$\hidewidth\cr}}
  {\ooalign{\raisebox{.14\height}{\scalebox{.7}{$\textstyle\sum$}}\cr\hidewidth$\textstyle\int$\hidewidth\cr}}
  {\ooalign{\raisebox{.2\height}{\scalebox{.6}{$\scriptstyle\sum$}}\cr$\scriptstyle\int$\cr}}
  {\ooalign{\raisebox{.2\height}{\scalebox{.6}{$\scriptstyle\sum$}}\cr$\scriptstyle\int$\cr}}
}

\newcommand{\backwardoperatorX}{\mathcal{L}_B^{(X)}}
\newcommand{\backwardoperatorY}{\mathcal{L}_B^{(Y)}}
\newcommand{\backwardoperatorI}{\mathcal{L}_B^{(i)}}

\newcommand{\evecy}{u^{(Y)}}
\newcommand{\tildeevecy}{\tilde{u}^{(Y)}}
\newcommand{\eveci}{u^{(i)}}
\newcommand{\tildeeveci}{\tilde{u}^{(i)}}

\newcommand{\pipm}{\pi_{y_1}}
\newcommand{\Dx}{D}
\newcommand{\pipmone}{\pi^{(1)}_{y_1}}
\newcommand{\pipmtwo}{\pi^{(2)}_{y_1}}

\newcommand{\rhopmone}{\rho^{(1)}_{y_1}}
\newcommand{\rhopmtwo}{\rho^{(2)}_{y_1}}
\newcommand{\sigmapmone}{\sigma^{(1)}_{y_1}}
\newcommand{\sigmapmtwo}{\sigma^{(2)}_{y_1}}
\newcommand{\czero}{c^{(y_1)}_0}
\newcommand{\cone}{c^{(y_1)}_1}
\newcommand{\ctwo}{c^{(y_1)}_2}
\newcommand{\cthree}{c^{(y_1)}_3}
\newcommand{\dzero}{d^{(y_1)}_0}
\newcommand{\done}{d^{(y_1)}_1}
\newcommand{\dtwo}{d^{(y_1)}_2}
\newcommand{\dthree}{d^{(y_1)}_3}

\newcommand{\pisigma}{\pi^{\sigma}}
\newcommand{\pisigmar}{\pi^{\sigma}_R}

\renewcommand{\contentsname}{}

\begin{document}

\title[Hidden state inference in drift-diffusive processes...]{Conditional splitting probabilities for hidden-state inference in drift-diffusive processes}

\author{Emir Sezik$^1$, Jacob Knight$^1$, Henry Alston$^2$, Connor Roberts$^1$, Thibault Bertrand$^1$, Gunnar Pruessner$^1$, Luca Cocconi$^3$}

\address{$^1$Department of Mathematics, Imperial College London, UK\\
$^2$Laboratoire de Physique de l'Ecole Normale Supérieure (LPENS), Paris, France\\
$^3$Max Planck Insitute for Dynamics and Self-Organization, Göttingen, Germany}
\ead{luca.cocconi@ds.mpg.de}
\vspace{10pt}
\begin{indented}
\item[]\today
\end{indented}

\begin{abstract}

Splitting probabilities quantify the likelihood of particular outcomes out of a set of mutually-exclusive possibilities for stochastic processes and play a central role in first-passage problems. For two-dimensional Markov processes $\{X(t),Y(t)\}_{t\in T}$, a joint analogue of the splitting probabilities can be defined, which captures the likelihood that the variable $X(t)$, having been initialised at $x_0 \in \mathbb{L}$, exits $\mathbb{L}$ for the first time via either of the interval boundaries \emph{and} that the variable $Y(t)$, initialised at $y_0$, is given by $y_{\rm exit}$ at the time of exit. 
We compute such joint splitting probabilities for two classes of processes: processes where $X(t)$ is Brownian motion and $Y(t)$ is a decoupled internal state, and unidirectionally coupled processes where $X(t)$ is drift-diffusive and depends on $Y(t)$, while $Y(t)$ evolves independently. 
For the first class we obtain generic expressions in terms of the eigensystem of the Fokker-Planck operator for the $Y$ dynamics, while for the second we carry out explicit derivations for three paradigmatic cases (run-and-tumble motion, diffusion in an intermittent piecewise-linear potential and diffusion with stochastic resetting). 
Drawing on Bayes' theorem, we subsequently introduce the related notion of conditional splitting probabilities, defined as the posterior likelihoods of the internal state $Y$ \emph{given} that the observable degree of freedom $X$ has undergone a specific exit event. After computing these conditional splitting probabilities, we propose a simple scheme that leverages them to partially infer the assumedly hidden state $Y(t)$ from point-wise detection events. 
\end{abstract}

\tableofcontents

\section{Introduction}\label{sec:intro}

Splitting probabilities express the likelihood of a specific outcome among a set of mutually exclusive possibilities for a stochastic process \cite{redner2001guide,van1992stochastic,hughes1996random,gardiner1985handbook}. For one-dimensional continuous Markov processes $\{X(t)\}_{t \in T}$ with $X(t)\in \mathbb{R}$ and time $t \in \mathbb{R}^{+}$ initialised at time $t_0$ and position $x_0$ in some interval $\mathbb{L} =[x_{\rm min},x_{\rm max}]$, splitting probabilities are associated with the complementary probabilities that the process exits the interval by crossing $x_{\rm min}$ before $x_{\rm max}$, or vice versa. Classical applications include variants of the Gambler's ruin problem \cite{redner2001guide}, where $x_{\rm min}=0$ and betting is interrupted upon reaching a target wealth $x_{\rm max}$; 
population dynamics, where $x_{\rm max}$ is the population size and splitting probabilities quantify the likelihood of fixation versus disappearance of a mutation \cite{moran1962statistical}; 
the melting of heteropolymers where $x_{\rm max}$ is the length of the polymer and the splitting probability captures the probability of melting completely vs returning to a helix state \cite{oshanin2009helix}; the transmission probability through a scattering medium, where $x_{\rm min}$ and $x_{\rm max}$ denote the location of actual material interfaces in space \cite{klinger2022splitting,baudouin2014signatures}; and decision theory, particularly Wald’s sequential
probability ratio test, where splitting probabilities can be related to the reliability of the decision \cite{roldan2015decision}, e.g.\ in the context of cellular sensing \cite{Siggia2013, Desponds2020, Tohme2025}. 
Splitting probabilities also play a key role in the study of rare events, e.g.\ the stochastic transition between metastable states in chemical reactions, where they are sometimes referred to as (forward) \emph{committor} or \emph{capacitance} functions \cite{vanden2010transition,das2022direct}. The splitting probabilities as a function of the initial value $x_0$ are given by the solution to the stationary Kolmogorov (sometimes, Fokker-Planck) backward equation of the process with suitable boundary conditions \cite{van1992stochastic,risken1996fokker}. For the study of chemical reactions, where the system of interest is often multi-dimensional and potentially out of equilibrium, analytical solutions may not be available and a variety of computational methods
have been developed to obtain numerical approximations \cite{singh2023variational,kang2024computing,singh2025variational}. For a broad class of stochastic processes, splitting probabilities can also be related to the cumulative distribution of a dual process confined between two reflecting boundaries by leveraging the so-called Siegmund duality \cite{gueneau2024siegmund,gueneau2024relating}.

Traditionally, determining the splitting probabilities has been framed as a question of the relative likelihood of achieving a ``desired'' outcome among a set of accessible possibilities. In the present work, for two-dimensional processes $\{X(t),Y(x)\}_{t \in T}$ where the outcomes are expressed in terms of the boundary points of an accessible interval $\mathbb{L}$ for the sub-process $\{X(t)\}$ only, we demonstrate that knowledge of the splitting probabilities provides also a tool for the Bayesian inference of the (assumedly hidden) \emph{internal} state $\{Y(t)\}$ at the time of exit from said interval. To the best of our knowledge, this kind of boundary-point inference was applied for the first time in Ref.~\cite{cocconi2023optimal} to the optimisation of power extraction from active particles with hidden states: while the self-propulsion force $Y(t)$ was assumed to be a zero-mean stochastic process, the sign of its conditional expectation was shown to depend on the point of exit, allowing an external observer to systematically apply forces opposite to the active one.
Implicit expressions for these so-called \emph{conditional} splitting probabilities have also recently been derived for one-dimensional run-and-tumble motion on piecewise-linear potentials \cite{roberts2023run}. 

In the following, we summarise these earlier findings alongside some new and more general results, with a particular focus on drift-diffusive processes with either discrete or continuous internal-state dynamics. 
After a mathematical introduction (Sec.~\ref{s:formalism}), we start by determining the conditional splitting probabilities for a process in which $X$ is a simple Brownian motion and the dynamics of $X$ and $Y$ are fully decoupled (Sec.~\ref{s:decoupled}), before considering more complicated cases where the $X$ dynamics are $Y$-dependent, but not vice versa (Sec.~\ref{s:coupled}). The latter is a common occurrence in minimal models of motile active matter, where, for instance, a fluctuating internal state can control the magnitude and orientation of the self-propulsion force  \cite{bechinger2016active,solon2015active}, as well as in simple descriptions of passive Brownian motion in nonequilibrium environments, such as fluctuating and time-dependent potentials \cite{alston2022non,cocconi2024ou2,santra2021brownian}.
In Sec. \ref{s:conclusion}, we summarise our results and comment on the applicability of this inference scheme. 

\section{Joint and conditional splitting probabilities}\label{s:formalism}

We consider a two-dimensional Markov process $\{X(t),Y(t)\}_{t \in T}$ characterised by state variables $X$ and $Y$. Throughout this work, we assume that the variable $X(t) \in \mathbb{R}$ is real and continuous, while $Y(t) \in \mathbb{Y}$ is real but may not be continuous. The Fokker-Planck equation capturing the time evolution of the joint probability density $P(X,Y;t)$ can be written generically as
\begin{equation}\label{eq:forward_FP}
    \partial_t P(X,Y;t) = \mathcal{L}_{F}^{(X)} P(X,Y;t) + \mathcal{L}_{F}^{(Y)} P(X,Y;t)
\end{equation}
with the initial condition $P(X,Y;t_0)=\delta(X-x_0)\delta(Y-y_0)$, where $\mathcal{L}^{(X)}$ and $\mathcal{L}_{F}^{(Y)}$ are the forward Fokker-Planck operators associated with the time evolution for the variables $X$ and $Y$, respectively \cite{risken1996fokker}. If $X(t)$ is a continuous drift-diffusive process, we may generally write $\mathcal{L}_{F}^{(X)}q \equiv \partial_X [D(X,Y)\partial_X q - W(X,Y)q]$ 
to allow both stochastic and deterministic contributions to depend on the current value of the \emph{internal} state $Y$, which we assume to be hidden to an external observer. 

We are interested in a general expression for a particular kind of splitting probability, namely the joint probability that the drift-diffusive sub-process $X(t)$, having been initialised at $x_0 \in \mathbb{L} = (x_{\rm min},x_{\rm max})$, exits $\mathbb{L}$ \emph{for the first time} either via $x_{\rm min}$ or $x_{\rm max}$ \emph{and} that the internal state, initialised at $y_0$, is given by $y_{\rm exit}$ at the time of exit. We denote this joint probability by $\Pi(x_{\rm exit},y_{\rm exit}|x_0,y_0)$ with $x_{\rm exit} \in \{x_{\rm min},x_{\rm max}\}$. It satisfies the stationary Kolmogorov backward equation \cite{redner2001guide,gardiner1985handbook}
\begin{equation}\label{eq:stat_kolm_back}
    0 = \mathcal{L}^{(X)}_B \Pi(x_{\rm exit},y_{\rm exit}|x_0,y_0) + \mathcal{L}^{(Y)}_B \Pi(x_{\rm exit},y_{\rm exit}|x_0,y_0),
\end{equation}
where $\mathcal{L}^{(X)}_B$ and $\mathcal{L}^{(Y)}_B$ are the backward Fokker-Planck operators, acting on $x_0$ and $y_0$, respectively. These are the adjoints of the operators appearing in Eq.~\eqref{eq:forward_FP}, i.e.\ $\mathcal{L}^{(i)}_B = (\mathcal{L}^{(i)}_{F})^\dagger$ with $i \in \{X,Y\}$, where they acted on the variables $X$ and $Y$, respectively. 
The joint probability is normalised,
\begin{equation}
   \SumInt_{x_{\rm exit} \in \mathbb{L},y_{\rm exit}\in\mathbb{Y}} \Pi(x_{\rm exit},y_{\rm exit}|x_0,y_0) = 1
\end{equation}
reflecting the fact that the particle will eventually leave the interval $\mathbb{L}$.

For fully permeable boundaries, where the particle's passage through the boundaries is not obstructed, the backward equation is complemented by the boundary conditions
\begin{subequations} \label{eq:conditions}
\begin{align}
    \Pi(x_{\rm exit},y_{\rm exit}| x_{\rm exit},y_0) &= \delta_{y_{\rm exit},y_0} \label{eq:cond_same}~,\\
    \Pi(x_{\rm exit},y_{\rm exit}|\neg x_{\rm exit},y_0)  &= 0~,\label{eq:cond_diff}
\end{align}    
\end{subequations}
where $ \delta_{y_1,y_2}$ is the Dirac delta function if $Y$ is continuous and the Kronecker delta if $Y$ is discrete, while $\neg x_{\rm exit} = x_{\rm min}$ if $x_{\rm exit} = x_{\rm max}$, and vice versa. The interpretation of the conditions in Eq. \eqref{eq:conditions} is straightforward: if the process is initialised at one of the boundaries, the exit condition is immediately satisfied and the exit value of the hidden process $\{Y(t)\}$ is fixed by the corresponding initialisation $y_0$. For partially permeable boundaries (e.g.\ due to stochastic switching between perfectly permeable and reflecting \cite{lawley2015new}), Robin boundary conditions may be used instead. In the simplest case where only one boundary is partially permeable (say $x_{\rm max}$), the boundary conditions become 
\begin{subequations}\label{eq:permeable_bc}
\begin{gather}
    \Pi(x_{\rm min}, y_{\rm exit}| x_{\rm min}, y_0) = \delta_{y_{\rm exit}, y_0}~, \\
    \left. \partial_x \Pi(x_{\rm min},y_{\rm exit}|x,y_0) \right|_{x=x_{\rm max}} + \kappa \Pi(x_{\rm min},y_{\rm exit}| x_{\rm max},y_0) = 0~, \\
     \Pi(x_{\rm max}, y_{\rm exit}| x_{\rm min}, y_0) = 0~, \\
     \left. \partial_x \Pi(x_{\rm max},y_{\rm exit}|x,y_0) \right|_{x=x_{\rm max}} + \kappa \Pi(x_{\rm max},y_{\rm exit}| x_{\rm max},y_0) =  \kappa \delta_{y_{\rm exit}, y_0}~,
\end{gather}
\end{subequations}
with $\kappa \geq 0$ controlling the permeability such that $\kappa = 0$ renders the boundary impermeable (reflecting) and $\kappa \rightarrow \infty$ fully permeable. The latter can be seen by dividing Eq. \eqref{eq:permeable_bc} through by $\kappa$ and taking the limit $\kappa \to \infty$ which imposes the absorbing boundary conditions in Eq. \eqref{eq:conditions}. For a one-dimensional process $\{X(t)\}_{t \in T}$, the marginal splitting probabilities, $\Pi(x_{\rm exit}|x)$, in the impermeable limit $\kappa \to 0$ are trivial, since exit can occur only via $x_{\rm min}$. However, the same is not true for the joint probability $\Pi(x_{\rm min},y_{\rm exit}|x_0,y_0)$, which will in general retain a non-trivial dependence on $y_{\rm exit}$.

Through the application of Bayes' theorem, the joint probability $\Pi(x_{\rm exit},y_{\rm exit}|x_0,y_0)$ can be combined with the prior $P(x_{\rm exit}|x_0,y_0) \equiv \SumInt_{y_{\rm exit}\in\mathbb{Y}} \Pi(x_{\rm exit},y_{\rm exit}|x_0,y_0)$, i.e.\ the marginal splitting probability of the drift-diffusive process alone, to obtain the \emph{conditional} splitting probability
\begin{equation}\label{eq:cond_split_def1}
    \Pi(y_{\rm exit}| x_{\rm exit}; x_0,y_0) = \frac{\Pi(x_{\rm exit},y_{\rm exit}|x_0,y_0)}{P(x_{\rm exit}|x_0,y_0)},
\end{equation}
which conveys the likelihood that the process, having been initialised at $(x_0,y_0)$, exits the interval $\mathbb{L}$ with internal state $y_{\rm exit}$ given that exit occurred through the boundary at $x_{\rm exit}$. When the initialisation is sampled from the joint prior $P_{\rm init}(x_0,y_0)$, the conditional splitting probability reads 
\begin{equation}\label{eq:cond_split_def2}
    \Pi(y_{\rm exit}| x_{\rm exit}) = \frac{\SumInt_{x_0 \in \mathbb{L},y_0\in\mathbb{Y}} \Pi(x_{\rm exit},y_{\rm exit}|x_0,y_0) P_{\rm init}(x_0,y_0)}{\SumInt_{x_0 \in \mathbb{L},y_0\in\mathbb{Y}} P(x_{\rm exit}|x_0,y_0) P_{\rm init}(x_0,y_0)}~.
\end{equation}
Assuming that either the initial condition or its distribution are known, Eqs.~\eqref{eq:cond_split_def1} and \eqref{eq:cond_split_def2} thus provide a tool to infer the hidden internal state based on the point-wise observation of an exit event. The observation itself may be triggered, in an experimental setting, by a pair of sensors located at the boundaries \cite{cocconi2023optimal}. In the impermeable limit $\kappa \to 0$, Eq.~\eqref{eq:cond_split_def2} becomes trivial since $X(t)$ exits the interval $\mathbb{L}$ from the permeable boundary with probability one, $\Pi(y_{\rm exit}|x_{\rm exit}) = \Pi(y_{\rm exit})$.

\section{Brownian motion with a decoupled hidden state}\label{s:decoupled}

We now consider processes $\{X(t),Y(t)\}_{t \in T}$ which are dynamically decoupled. Equivalently, the Fokker-Planck operators satisfy $\mathcal{L}^{(i)}_{B} = \mathcal{L}^{(i)}_{B}(i,\partial_i)$ for $i \in \{X,Y\}$, such that $\mathcal{L}^{(i)}_{B}$ acts and depends only on the variable $i$. The set of eigenvectors of these operators is guaranteed to form an eigenbasis by Sturm-Liouville theory \cite{risken1996fokker}. Specifically, if $\backwardoperatorI$ possesses a discrete non-positive spectrum with $\eveci_n$ its $n^{\rm th}$ eigenfunction, i.e.\ $\backwardoperatorI \eveci_n = -\lambda_n^{(i)} \eveci_n$, the following relations hold
\begin{subequations} \label{eq:orhonormalandcomplete}
    \begin{gather}
        \langle \tildeeveci_n, \eveci_m \rangle = \delta_{n,m}~, \\
        \sum_n \tildeeveci_n(x) \eveci_n(x') = \delta_{x,x'}~,
    \end{gather}
\end{subequations}
where $\tildeeveci_n$ is the $n^{\rm th}$ eigenfunction of $\mathcal{L}_{F}^{(i)}$ and $ \langle \cdot, \cdot \rangle$ is an appropriate inner product in the space on which $\mathcal{L}_{B,F}^{(i)}$ acts with the weight attached to $\tildeeveci$. 
If the spectrum of $\backwardoperatorI$ is continuous, these relations hold as long as the summations are replaced with integrals. 
If the operators are not self-adjoint, $\mathcal{L}^{(i)}_{B}\neq \mathcal{L}^{(i)}_{F}$, the set of eigenfunctions $\{\eveci\}$ and $\{\tildeeveci\}$ are not the same.  

Parametrising the initial conditions as $x,y$ for notational ease and introducing the shorthand $\pi(x,y)\equiv \Pi(x_{\rm exit},y_{\rm exit}|x,y)$,
we can express $\pi(x,y)$ in terms of the eigenbasis of $\backwardoperatorY$ as
\begin{equation}\label{eq:exp_in_basis}
    \pi(x,y) = \sum_{n = 0}^{\infty} \pi_{n}(x) \evecy_{n}(y) ~.
\end{equation}
For fully permeable boundaries, Eq.~\eqref{eq:conditions}, using the orthogonality and completeness properties of $\evecy_n(y)$, one can derive boundary conditions on $\pi_{n}(x)$:
\begin{subequations} \label{eq:bcs_projected}
\begin{align}
    \pi_{n}(x_{\rm exit}) &= \tildeevecy_n(y_{\rm exit}) \\
    \pi_{n}(\neg x_{\rm exit}) &= 0~.
\end{align}
\end{subequations}
For the case of a partially permeable boundary at $x_{\rm max}$, Eq.~\eqref{eq:permeable_bc}, one has instead the following boundary conditions:
\begin{subequations} \label{eq:bcs_perm_projected}
    \begin{gather}
        \pi_{n}(x_{\rm min}) =
        \begin{cases}
            \tildeevecy_n(y_{\rm exit}) &\text{if}\quad  ~x_{\rm exit} = x_{\rm min} \\
            0 &\text{if}\quad ~x_{\rm exit}  = x_{\rm max}
        \end{cases}
        \\
        \left. \partial_x\pi_{n}(x) \right.|_{x = x_{\rm max}} =
        \begin{cases}
            -\kappa \pi_{n}( x_{\rm max}) &\text{if}\quad  ~x_{\rm exit} = x_{\rm min} \\
            -\kappa \pi_{n}( x_{\rm max}) + \kappa \tildeevecy_n(y_{\rm exit}) &\text{if}\quad ~x_{\rm exit}  = x_{\rm max}
        \end{cases}
        %
    \end{gather}
\end{subequations}
where we have used $\pi_n(x) = \int dy ~\tilde{u}^{(Y)}_n(y) \pi(x,y)$ which can be obtained by integrating Eq. \eqref{eq:exp_in_basis} over $\tilde{u}^{(Y)}_n(y)$ and using the orthogonality relations in Eq. \eqref{eq:orhonormalandcomplete}. Plugging Eq.~\eqref{eq:exp_in_basis} back into Eq.~\eqref{eq:stat_kolm_back} and taking the inner product $\langle \tildeevecy_n, \cdot\rangle$, we additionally get 
\begin{equation}\label{eq:reduced_back_fp}
    \backwardoperatorX \pi_{n}(x) = \lambda_n^{(Y)} \pi_n(x)~,
\end{equation}
where $\mathrm{Re}[\lambda_n^{(Y)}] \geq 0$ due to the negative semi-definiteness of the Fokker-Planck operators. This is an eigenvalue equation for $\pi_n(X)$ with eigenvalue $\lambda_n^{(Y)}$. Upon solving the ODE in Eq. \eqref{eq:reduced_back_fp} for a given process defined by the Fokker-Planck operator, one can construct the conditional splitting probability using Eq. \eqref{eq:exp_in_basis}. This, though straightforward, is not analytically possible for an arbitrary Markov process. To make analytical progress, we henceforth assume that $X(t)$ is a homogeneous Brownian motion, $\mathcal{L}_B^{(X)}=D\partial_{x}^2$. Then, Eq.~\eqref{eq:reduced_back_fp} can always be solved by the linear combination,
\begin{equation}\label{eq:superp_sol}
    \pi_n(x) = c_{n,1} \exp\left(\sqrt{\frac{\lambda_n^{(Y)}}{D}}x \right) + c_{n,2} \exp\left(-\sqrt{\frac{\lambda_n^{(Y)}}{D}}x \right) \equiv c_{n,1} f_{n,1}(x) + c_{n,2} f_{n,2}(x)~.
\end{equation}
Applying Eq.~\eqref{eq:bcs_projected}, the constants of integration are fixed as
\begin{subequations} \label{eq:constants_of_integration}
\begin{gather}
    c_{n,1} = \tildeevecy_n(y_{\rm exit})\frac{f_{n,2}(\neg x_{\rm exit})}{f_{n,1}(x_{\rm exit})f_{n,2}(\neg x_{\rm exit}) - f_{n,1}(\neg x_{\rm exit})f_{n,2}(x_{\rm exit})} ~,\\
    c_{n,2} = - \tildeevecy_n(y_{\rm exit})\frac{f_{n,1}( \neg x_{\rm exit})}{f_{n,1}(x_{\rm exit})f_{n,2}(\neg x_{\rm exit}) - f_{n,1}(\neg x_{\rm exit})f_{n,2}(x_{\rm exit})}~.
\end{gather}
\end{subequations}
Combining Eqs. \eqref{eq:exp_in_basis}, \eqref{eq:superp_sol} and \eqref{eq:constants_of_integration}, we arrive at the general expression 
\begin{align}
    &\Pi(x_{\rm exit},y_{\rm exit}|x,y) = \nonumber \\
    &\SumInt_n \frac{\tildeevecy_n(y_{\rm exit})\evecy_n(y) \left[ f_{n,2}(\neg x_{\rm exit}) f_{n,1}(x) -  f_{n,1}( \neg x_{\rm exit})f_{n,2}(x) \right]}{f_{n,2}(\neg x_{\rm exit})f_{n,1}(x_{\rm exit}) - f_{n,1}(\neg x_{\rm exit})f_{n,2}(x_{\rm exit})}~. \label{eq:gen_noint_f}
\end{align}
Without loss of generality, we shift $X(t)$ such that $x_{\rm min}=-L/2$ and $x_{\rm max}=L/2$. Then, the exponential factors in Eq.~\eqref{eq:gen_noint_f} for $x_{\rm exit} = \sigma L/2$
may be written more compactly as 
\begin{equation} \label{eq:gen_noint_sinh}
    \Pi(\sigma L/2,y_{\rm exit}|x,y) = \SumInt_n \frac{\tildeevecy_n(y_{\rm exit})\evecy_n(y)}{\sinh\left(\sqrt{\frac{\lambda^{(Y)}_n}{D}} L \right)} \sinh\left[\sqrt{\frac{\lambda^{(Y)}_n}{D}} \left( \frac{L}{2} +\sigma x \right) \right]~,
\end{equation}
where $\sigma \in \{-1,1\}$ and the $n=0$ contribution,where $\lambda_{n=0}^{(Y)} = 0$, is understood to be evaluated as the limit $\lambda_0 \to 0$.
Again, this holds when $X(t)$ is a homogeneous diffusive process and otherwise assumes only the orthogonality and completeness relations in Eq.~\eqref{eq:orhonormalandcomplete}. 
For a partially permeable Robin boundary condition at $x_{\rm max} = L/2$, Eq.~\eqref{eq:permeable_bc}, a similar derivation can be performed to obtain the splitting probability of exit from $x = \sigma L/2$ as 
\begin{small}
\begin{subequations}\label{eq:joint_split_prob_semiperm_pair}
\begin{align}
    &\Pi(-L/2,y_{\rm exit}|x,y) = \SumInt_n \frac{\tildeevecy_n(y_{\rm exit})\evecy_n(y)}{ \sqrt{\frac{\lambda^{(Y)}_n}{D}}\cosh\left(\sqrt{\frac{\lambda^{(Y)}_n}{D}} L \right)  + \kappa \sinh\left(\sqrt{\frac{\lambda^{(Y)}_n}{D}} L \right)}  \nonumber \notag \\
    & \qquad \qquad\qquad\qquad \times \left[ \sqrt{\frac{\lambda^{(Y)}_n}{D}}\cosh\left(\sqrt{\frac{\lambda^{(Y)}_n}{D}} \left( \frac{L}{2}  - x \right) \right) +\kappa \sinh\left(\sqrt{\frac{\lambda^{(Y)}_n}{D}} \left( \frac{L}{2}  -x \right) \right) \right]~, \label{eq:joint_split_prob_semiperm_left} \\
    &\Pi(L/2,y_{\rm exit}|x,y) = \SumInt_n \frac{\tildeevecy_n(y_{\rm exit})\evecy_n(y)}{ \sqrt{\frac{\lambda^{(Y)}_n}{D}}\cosh\left(\sqrt{\frac{\lambda^{(Y)}_n}{D}} L \right)  + \kappa \sinh\left(\sqrt{\frac{\lambda^{(Y)}_n}{D}} L \right)} \kappa \sinh\left(\sqrt{\frac{\lambda^{(Y)}_n}{D}} \left( \frac{L}{2} + x \right) \right) ~,  \label{eq:joint_split_prob_semiperm_right}
\end{align}
\end{subequations}
\end{small}
which recovers Eq.~\eqref{eq:gen_noint_sinh} in the limit $\kappa \to \infty$, as expected. Either of these Eqs.~\eqref{eq:gen_noint_sinh}--\eqref{eq:joint_split_prob_semiperm_pair} may be combined with Eq.~\eqref{eq:cond_split_def2} to derive the associated conditional splitting probabilities. The equivalent form for the partially permeable boundary at $-L/2$ is obtained by symmetry arguments.

Taking the limit of vanishing diffusion constant for the simple Brownian motion, $D \to 0$, we observe that all contributions to Eqs.~\eqref{eq:joint_split_prob_semiperm_left} and \eqref{eq:joint_split_prob_semiperm_right} other than that associated to the smallest eigenvalue $\lambda_0^{(Y)}=0$ (corresponding to the unique stationary state $\tildeevecy_0(y_{\rm exit})\equiv \rho^{(Y)}_{\rm steady}(y_{\rm exit})$ of the hidden internal process) vanish exponentially. Combining this with $\evecy_0(y_{\rm exit})  = 1$, which follows from normalisation of $\rho^{(Y)}_{\rm steady}$ and Eqs.~(\ref{eq:orhonormalandcomplete}), we conclude that the joint splitting probability in this limit is independent of $y$ and factorises as
\begin{subequations}
\begin{align}\label{eq:decoupled_factorisation}
    \lim_{D \to 0} \Pi( - L/2,y_{\rm exit}|x,y)  = \rho^{(Y)}_{\rm steady}(y_{\rm exit}) \mathcal{S}(x) \\
    \lim_{D \to 0} \Pi( L/2,y_{\rm exit}|x,y)  = \rho^{(Y)}_{\rm steady}(y_{\rm exit}) \tilde{\mathcal{S}}(x) 
\end{align}
\end{subequations}
where $\mathcal{S}(x) \equiv (L/2 + \kappa^{-1} - x)/(L+\kappa^{-1})$ and $\tilde{\mathcal{S}}(x) = (L/2 + x)/(L + \kappa^{-1})$ are the (complementary) marginal splitting probabilities for simple Brownian motion with a partially permeable boundary condition at $x_{\rm max}=  L/2$. This is rationalised physically by noticing that an exit event takes long enough for the internal dynamics to relax to its steady state by the time of exit. Since Eqs.~\eqref{eq:gen_noint_sinh}--\eqref{eq:joint_split_prob_semiperm_right} are functions of $\lambda_n^{(Y)}L^2/D$, taking the $D \to 0$ limit is physically equivalent to taking the $L \to \infty$ limit.

A similar factorisation occurs if we average the conditional splitting probabilities \eqref{eq:joint_split_prob_semiperm_left} and \eqref{eq:joint_split_prob_semiperm_right} over the initial internal state $y$ using the stationary probability of the internal process, $P_{\rm init}(Y) = \rho^{(Y)}_{\rm steady}(Y)$.
Such factorisations indicate that, in the absence of coupling, inference of the internal state is possible only in the transient regime and upon a non-stationary initialisation of $Y(t)$. 

Finally, marginalising Eq.~\eqref{eq:joint_split_prob_semiperm_left} (or \eqref{eq:joint_split_prob_semiperm_right}) by summing/integrating out $y_{\rm exit}$ returns the marginal splitting probability of simple Brownian motion $\mathcal{S}(X)$, as expected. To see this, one may use the fact that $\SumInt_{y_{\rm exit}} \tildeevecy_n(y_{\rm exit}) \equiv \langle \tildeevecy_n(y_{\rm exit}), \evecy_0(y_{\rm exit})\rangle= \delta_{n,0}$ and simplify the (otherwise ill-defined) remaining $n=0$ term by taking a formal limit $\lambda_0\to 0$. 

In the following, we apply the general results obtained thus far to two illustrative cases. For the sake of brevity, we shall focus solely on $\Pi( - L/2,y_{\rm exit}|x,y)$, the probability (density) to exit via the left boundary in state $y_{\rm exit}$, having started from $x$ in state $y$. For $\kappa \to \infty$, the second joint probability $\Pi( L/2,y_{\rm exit}|x,y)$ may be obtained by straightforward symmetry arguments, while for $\kappa \to 0$ it vanishes trivially.

\subsection{Ripening-spoiling process}\label{ss:rip_spoil}
To illustrate the non-trivial nature of the joint probability $\Pi(x_{\rm exit},y_{\rm exit}|x,y)$, we consider a Brownian particle that, once initialised, undergoes a process of ``ripening'' followed by ``spoiling''. Both transitions are characterised by irreversible Poisson jump processes with homogeneous rates $r$ (from unripe to ripe, $U \to R$) and $s$ (from ripe to spoiled, $R \to S$), respectively. The reaction chain is akin to that of a susceptible-infected-recovered (SIR) model 
\cite{kroger2020analytical}, or of a simple model of protein folding/binding and subsequent degradation. The forward Fokker-Planck operator $\mathcal{L}_F^{(Y)}$ is simply the transition-rate matrix: 
\begin{equation}
    \mathcal{L}_F^{(Y)} = 
    \begin{pmatrix}
        -r & 0 & 0 \\
        r & -s & 0 \\
        0 & s & 0
    \end{pmatrix}
    ~,
\end{equation} 
from which we can directly obtain the eigensystem by finding the right eigenvectors of $\mathcal{L}_F^{(Y)}$ and $\mathcal{L}_B^{(Y)} $ ($= (\mathcal{L}_F^{(Y)})^{T}$):
\begin{subequations}
\begin{align}
    \lambda^{(Y)}_0 = 0, \qquad&\tildeevecy_0 = (0,0,1), &&\evecy_0=(1,1,1)~,  \\
    \lambda^{(Y)}_1 = r, \qquad&\tildeevecy_1 = (s-r,r,-s), &&\evecy_1=\left(\frac{1}{s-r},0,0\right)~, \\
    \lambda^{(Y)}_2 = s,  \qquad&\tildeevecy_2 = (0,-1,1), &&\evecy_2=\left(-\frac{r}{r-s},-1,0\right)~,
\end{align}
\end{subequations}
with vector components in the order $U,R,S$. The eigenvectors satisfy Eq.~\eqref{eq:orhonormalandcomplete} with $\langle.,.\rangle$ the usual dot product.
In the following, we assume for simplicity that $r \neq s$ as otherwise the eigenvalues are degenerate and the eigenvectors do not span the vector space. 
Using Eq.~\eqref{eq:joint_split_prob_semiperm_left} with $y=U$ to indicate initialisation in the unripe state, we obtain the following expression for the probability that the particle exits the left boundary while in a ripe (R) state:
\begin{align}
    &\Pi(-L/2,R|x,U) = 
    \sum_{n \in \{0,1,2\}} \frac{\tildeevecy_{n,R} \evecy_{n,U} \left[\kappa \sinh\left(\chi^{(Y)}_n\left(\frac{1}{2}-\frac{x}{L}\right)\right) + \frac{\chi^{(Y)}_n}{L}\cosh\left(\chi^{(Y)}_n\left(\frac{1}{2}-\frac{x}{L}\right)\right)\right]}{\kappa \sinh\left(\chi^{(Y)}_n\right) + \frac{\chi^{(Y)}_n}{L}\cosh\left(\chi^{(Y)}_n \right)}~,
    \label{eq:joinprob_ripening}
\end{align}
where $\chi^{(Y)}_n = L\sqrt{\lambda_n/D}$ and the subscripts $R,U$ denote the corresponding vector components. 

\begin{figure}
    \centering
    \includegraphics[width=\linewidth]{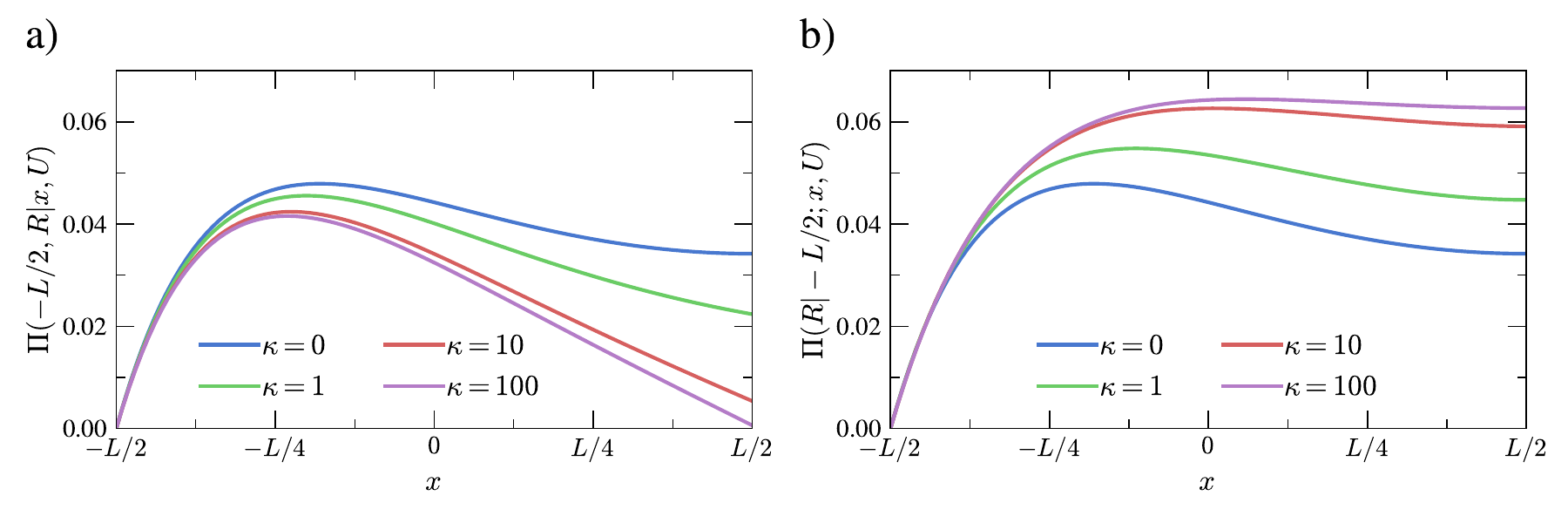}
    \caption{ Splitting probabilities for the ripening-spoiling process. a): Joint probability that a diffusive particle initialised at $x$ in the unripe (U) state leaves from the boundary at $-L/2$ and in a ripe (R) state, as given by Eq.~\eqref{eq:joinprob_ripening}. Lines correspond to different values of the permeability $\kappa$ of the right boundary. b): Conditional splitting probability that the particle leaves in state R, \textit{given} that it has left the interval at $-L/2$, obtained using equation Eq.~\eqref{eq:cond_split_def1}. 
    The parameter values used in this figure are $r=1, s=10, D=0.3$ and $L=1$. }
    \label{fig:ripen_example}
\end{figure}

The joint probability \eqref{eq:joinprob_ripening} is plotted as a function of the initial position $X$ and for different values of the permeability $\kappa$ in Fig.~\ref{fig:ripen_example}a. We observe that the probability of exiting from the boundary at $x_{\rm exit}=-L/2$ while in a ripe state exhibits a non-monotonic dependence on the initial $x$ and remains non-trivial even in the presence of a perfectly reflective boundary ($\kappa=0$) at $x_{\rm exit}=L/2$. In particular, as $x\to-L/2$, the particle becomes more likely to cross the left boundary before ripening and the conditional splitting probability tends to zero. For $x\to L/2$ and in the presence of a fully permeable boundary ($\kappa \to \infty$), the probability also vanishes as the particle is increasingly likely to escape from the right boundary or spoil as it reaches $x_{\rm exit}$. As the permeability of the right boundary decreases (decreasing $\kappa$), a particle initialised in its proximity has a higher chance to reach the left boundary before escaping via the right boundary and the joint splitting probability increases relatively for all $x$. 
In general, since ripe particles spoil after a characteristic waiting time $s^{-1}$, only trajectories of duration $\tau$ in the range $r^{-1}\lesssim \tau \lesssim r^{-1}+s^{-1}$ contribute to the splitting probabilities for exit in the ripe state. This effect is at the basis of the non-monotonic dependence on $x$ (more prominent as $L$ increases) featuring in the conditional splitting probabilities for all values of $\kappa$, Fig.~\ref{fig:ripen_example}b.

\subsection{Ornstein-Uhlenbeck process}\label{ss:ou}
\begin{figure}
    \centering
    \includegraphics[width=0.6\linewidth]{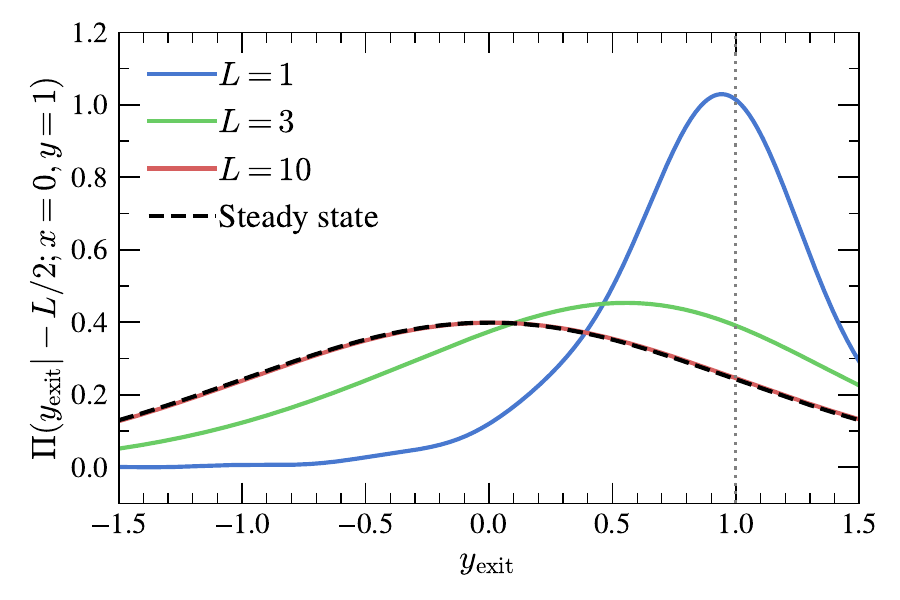}
    \caption{Conditional splitting probability density that a particle with an internal state undergoing an OU process will exit the interval with internal state $y_{\mathrm{exit}}$  given that it left through the boundary at $-L/2$, shown for different values of $L$. 
    The particle is initialised at $x=0$ with its internal OU process initialised at $y=1$. This probability is obtained by applying Eq.~\eqref{eq:cond_split_def1} to Eq.~\eqref{eq:no_coupl_pi}. 
    The parameter values used in this figure are $D=1$, $D_Y=1$ and $\mu=1$.
    }
    \label{fig:hidden_OU}
\end{figure}

To provide an example where the internal variable is characterised by a continuous-state stochastic process,  we take the $Y(t)$ dynamics to be a zero-mean Ornstein-Uhlenbeck (OU) process, $\dot{Y}=-\mu Y + \sqrt{2D_Y} \eta$ with $\mu^{-1}$ the characteristic relaxation time and $\eta$ a zero-mean Gaussian white noise---a paradigmatic example of a continuous Markov process. The associated backward Fokker-Planck operator reads $\backwardoperatorY = D_Y \partial_Y^2 - \mu Y \partial_Y$. The left and right eigenfunctions of the backward operator can be expressed in terms of Hermite polynomials (in the probabilist convention) \cite{risken1996fokker,bothe2021doi}. Defining $\ell^2=D_Y/\mu$, the eigensystem is  ($n=0,1,2,...$) \cite{risken1996fokker}:
\begin{equation}
    \lambda^{(Y)}_n = n \mu,\quad \tildeevecy_{n}(Y) = \frac{H_n(Y/\ell)}{n! \ell \sqrt{2\pi}} e^{-Y^2/(2\ell^2)}, \quad \evecy_{n}(Y) = H_n(Y/\ell)~.
\end{equation}
Assuming fully permeable boundaries ($\kappa \to \infty$), Eq.~\eqref{eq:gen_noint_sinh} can then be used to obtain
\begin{align}
        \Pi(-L/2,y_{\rm exit}|x,y) = 
        \frac{e^{-y_{\rm exit}^2/(2\ell^2)}}{\sqrt{ 2\pi} \ell}\sum_{n = 0}^{\infty} \frac{H_n\left( \frac{y}{\ell}\right) H_n\left( \frac{y_{\rm exit}}{\ell}\right)}{n! \sinh\left(\sqrt{\frac{\lambda^{(Y)}_n}{D}} L\right)} \sinh \left(\sqrt{\frac{\lambda^{(Y)}_n}{D}}\left( \frac{L}{2} - x\right) \right) ~.
        \label{eq:no_coupl_pi}
\end{align}
The corresponding conditional splitting probability, obtained using equation Eq.~\eqref{eq:cond_split_def1}, is shown in Fig.~\ref{fig:hidden_OU} for different values of the domain size $L$. For small $L$, when the mean escape time from either boundary is small compared to the characteristic timescale $\mu^{-1}$ of the OU dynamics, the posterior likelihood is concentrated around the initial $y$. As $L$ increases, the particle spends more time before leaving the region $\mathbb{L}$ and the posterior likelihood converges to the steady-state distribution of the hidden OU process, up to a proportionality constant given by the marginal splitting probability of the diffusive process, see Eq.~\eqref{eq:decoupled_factorisation}.\\

\section{Drift-diffusion with unidirectional coupling to a hidden state}\label{s:coupled}

Another class of models exhibiting non-trivial splitting probabilities are drift-diffusive processes $X(t)$ with a fluctuating internal state $Y(t)$ controlling the deterministic contribution to the dynamics. The process $\{Y(t)\}_{t\in T}$ itself is assumed to evolve in an $X$-independent fashion.  
This is a common occurrence in minimal models of motile active matter \cite{bechinger2016active,solon2015active} and, indeed, the concept of a \emph{conditional} splitting probability was first introduced in Ref.~\cite{cocconi2023optimal} as a tool to infer the internal state of run-and-tumble (RnT) particles \cite{solon2015active,garcia2021run} subject to an external control force. Another set of models belonging to this class are colloidal particles either in fluctuating external potentials \cite{roberts2023run,alston2022non,cocconi2024ou2,santra2021brownian,roberts2024ratchet}
or interacting via fluctuating pair-potentials \cite{cocconi2023active,alston2022intermittent}. 

In the presence of unidirectional coupling between the state variables, the derivation of the conditional splitting probabilities, as defined in Sec.~\ref{s:formalism}, becomes considerably more involved. Here, we carry out this calculation on three paradigmatic cases of the $X$-dynamics: free RnT motion (drawing on \cite{cocconi2023optimal}), thus generalising the known (unconditional) splitting probabilities for a free RnT particle \cite{malakar2018steady}, Brownian motion in an intermittent piecewise-linear potential (a flashing ratchet) \cite{roberts2024ratchet, astumian1994, astumian2002} and Brownian motion with stochastic resetting \cite{switkes2004unbiased,evans2011diffusion,evans2020stochastic}. Unlike the decoupled case considered in Sec.~\ref{s:decoupled}, in these examples inference of the internal state remains possible even when the initial value of the hidden state is distributed according to its stationary marginal distribution (cf.~Eq.~\eqref{eq:decoupled_factorisation}).

\subsection{Run-and-tumble dynamics}\label{ss:RnT_split}

We consider an RnT process governed by the Langevin equation $\dot{X}(t) = \nu Y(t) + \sqrt{2D} \xi(t)$, where the normalised self-propulsion state $Y \in \{-1,1\}$ follows a symmetric telegraph process with switching rate $\alpha$ [Supp.~Sec.~SVI in Ref.~\cite{cocconi2023optimal}].
We say the particle is in a right-moving (respectively, left-moving) state at time $t$ if $Y(t) = +1$ (respectively, $Y(t)=-1$).
In the following, we will use $ \Pi(\sigma L/2,y_{\rm exit}|x_0,y_0)$ to denote the joint splitting probability with $\sigma, y_0, y_{\rm exit} \in \{-1,+1\}$. Due to symmetry considerations, we need to compute only four of the eight possible combinations of parameters; the rest are then obtained using the following relation: 
\begin{equation}
    \Pi(\sigma L/2, y_{\rm exit}| x_0, y_0) = \Pi(-\sigma L/2, -y_{\rm exit}| -x_0, -y_0)~.
\end{equation}
As such, we will calculate only the splitting probabilities to exit at the left-hand boundary, $x = -L/2$. To ease notation, we thus define $\pi_{y_{\rm exit}}(x_0, y_0) \equiv \Pi_{y_{\rm exit}}^{ \sigma = -1}(x_0, y_0)$.

\begin{figure}
\includegraphics[width=\textwidth]{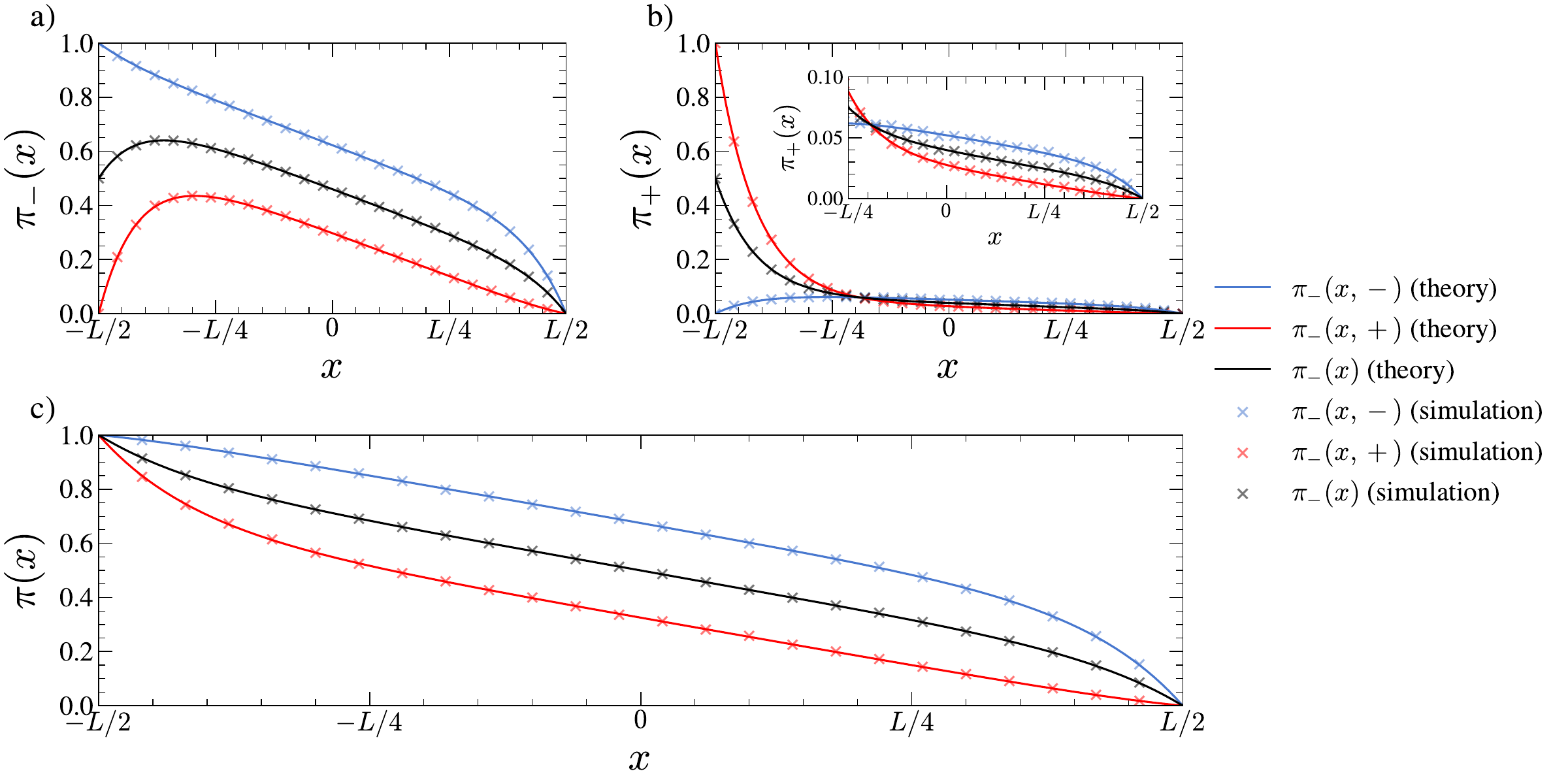}%
\caption{ Joint splitting probabilities, $\pi_{y_1}(x_0)$, as a function of initialisation position $x_{0}$, for an RnT particle with $\nu=1$, $D = 0.1$ and $\alpha = 2$, to cross the left-hand boundary at $x_{\rm min}=-L/2$, before crossing the right-hand boundary at $x=L/2$, if it exits as (a) a left mover, (b) a right mover, or (c) either a left mover or a right mover. In each subfigure, red (respectively, blue/black) lines indicate the initialisation of a right mover (respectively, left mover/equal superposition of a left mover and a right mover). In the inset of subfigure (b) the joint splitting probabilities are plotted for restricted range $x \in [-L/4, L/2]$ to distinguish the magnitude of different quantities. Subfigure (c) corresponds to the (unconditional) splitting probabilities first derived in Ref.~\cite{malakar2018steady}. Monte-Carlo simulations (markers) were performed by numerically integrating the Langevin equation, $\dot{X} = \nu Y(t) + \sqrt{2D_{X}}\xi(t)$, in timesteps of $\Delta t = 10^{-5}$ to determine the proportion of times the particle exits through the left-hand boundary for $10^{5}$ realisations at each $x_{0}/L = 0, 0.01, \dots, 1$. Figure adapted from Ref.~\cite{cocconi2023optimal}. }\label{fig:SplittingProbabilities}
\end{figure}

The Kolmogorov backward equation \eqref{eq:stat_kolm_back} amounts to two sets of coupled ODEs. Denoting $y_{\rm exit} = y_1 \in \{-1,1\}$, we find
\begin{subequations}\label{app:eq:SplittingProbCoupledODEs}
    \begin{alignat}{2}
        0 &= \nu \frac{\mathrm{d}\pi_{y_1}(x,+)}{\mathrm{d}x} + D \frac{\mathrm{d}^{2}\pi_{y_1}(x,+)}{\mathrm{d}x^{2}} + \alpha\left(\pi_{y_1}(x,-) - \pi_{y_1}(x,+)\right) \label{app:eq:SplittingProbCoupledODEsRightMover}~,\\
        0 &= -\nu\frac{\mathrm{d}\pi_{y_1}(x,-)}{\mathrm{d}x} + D \frac{\mathrm{d}^{2}\pi_{y_1}(x,-)}{\mathrm{d}x^{2}} + \alpha\left(\pi_{y_1}(x,+) - \pi_{y_1}(x,-)\right) \label{app:eq:SplittingProbCoupledODEsLeftMover}~,
    \end{alignat}
\end{subequations}
Since $\pi_{+}(x,y_0)$ and $\pi_{-}(x,y_0)$ are governed by the same ODEs, we proceed to solve Eq.~(\ref{app:eq:SplittingProbCoupledODEs}) up until the point of applying the different boundary conditions for each case, which are (cf.~Eq.~\eqref{eq:conditions})
\begin{equation}\label{app:eq:SplitProbBoundaryConditions}
    \pi_{y_1}(-L/2,y_0) = \delta_{y_1, y_0}~, \quad
    \pi_{y_1}(L/2,y_0) = 0~, \quad
\end{equation}
where we assumed fully permeable boundaries ($\kappa \to \infty$).
Defining the linear superpositions $\rho_{y_1}(x) \equiv \pi_{y_1}(x,-) + \pi_{y_1}(x,+)$ and $\sigma_{y_1}(x) \equiv \pi_{y_1}(x,-) - \pi_{y_1}(x,+)$, we may recast Eq.~\eqref{app:eq:SplittingProbCoupledODEs} as
\begin{subequations}
    \begin{alignat}{2}
        0 &= \frac{\mathrm{d}^{2}\rho_{y_1}(x)}{\mathrm{d}x^{2}} - \frac{\nu}{D} \frac{\mathrm{d}\sigma_{y_1}(x)}{\mathrm{d}x} \label{app:eq:DensityODE}~,\\
        0 &= \frac{\mathrm{d}^{2}\sigma_{y_1}(x)}{\mathrm{d}x^{2}} - \frac{\nu}{D} \frac{\mathrm{d}\rho_{y_1}(x)}{\mathrm{d}x} -\frac{2\alpha}{D}\sigma_{y_1}(x) \label{app:eq:PolarityODE}~,
    \end{alignat}
\end{subequations}
with boundary conditions
\begin{equation}\label{app:eq:SplitProbDensityPolarityBoundaryConditions}
    \rho_{y_1}\left(-L/2\right) = 1~, \quad
    \sigma_{y_1}\left(-L/2\right) = -y_1~, \quad
    \rho_{y_1}\left(L/2\right) = 0~, \quad
    \sigma_{y_1}\left(L/2\right) = 0~. \quad
\end{equation}
Integrating Eq.~(\ref{app:eq:DensityODE}) and after some rearrangements, we eventually obtain
\begin{subequations}
\begin{align}\label{app:eq:SigmaRhoSolution}
     \sigma_{y_1}(x) &= c_{3}^{(y_1)}e^{kx} + c_{2}^{(y_1)}e^{-kx} + c_{1}^{(y_1)} \\
     \rho_{y_1}(x) &= \frac{\nu}{D k} \left( c_{4}^{(y_1)} + c_{3}^{(y_1)}e^{kx} - c_{2}^{(y_1)}e^{-kx}\right) - \frac{2\alpha}{\nu} c_{1}^{(y_1)} x ~.
\end{align}
\end{subequations}
where $k = \sqrt{\nu^{2}/D^{2} + 2\alpha/D} > 0$ and the $c_{i}^{(y_1)}$ are constants of integration that depend on $y_1$. 
Applying the boundary conditions,  Eq.~(\ref{app:eq:SplitProbDensityPolarityBoundaryConditions}), these constants are determined as
\begin{subequations}\label{app:eq:SplittingProbConstants}
    \begin{alignat}{4}
        c_{1}^{(y_1)} &= \frac{1}{2}\frac{\cosh(\frac{kL}{2}) -y_1 \frac{\nu}{D k}\sinh(\frac{kL}{2})}{\frac{\alpha L}{\nu}\cosh(\frac{kL}{2}) + \frac{\nu}{D k}\sinh(\frac{kL}{2})}~,\\
        c_{2}^{(y_1)} &= -\frac{y_1}{4}\left( \frac{\frac{\alpha L}{\nu}+ y_1 }{\frac{\alpha L}{\nu} \cosh(\frac{kL}{2}) + \frac{\nu}{D k}\sinh(\frac{kL}{2})} + \mathrm{csch}\left(\frac{kL}{2}\right) \right)~,\\
        c_{3}^{(y_1)} &= - \frac{y_1}{4}\left(\frac{\frac{\alpha L}{\nu} + y_1 }{\frac{\alpha L}{\nu} \cosh(\frac{kL}{2}) + \frac{\nu}{D k}\sinh(\frac{kL}{2})} - \mathrm{csch}\left(\frac{kL}{2}\right) \right)~,\\
        c_{4}^{(y_1)} &= \frac{1}{2} \left( \frac{D k}{\nu} -y_1 \coth\left(\frac{kL}{2}\right) \right)~.
    \end{alignat}
\end{subequations}
The joint splitting probabilities are then fully determined by combining Eqs.~(\ref{app:eq:SigmaRhoSolution})--(\ref{app:eq:SplittingProbConstants}) with the definitions of $\pi_{y_1}(x,y_0)$ in terms of $\rho_{y_1}(x)$ and $\sigma_{y_1}(x)$.
They are plotted in Fig.~\ref{fig:SplittingProbabilities}, where we also compare our analytical results with Monte-Carlo simulations as a sanity check. Interestingly, we observe that there can be a higher likelihood
to observe a right mover at the left-hand boundary if it is first initialised as a left mover (Fig.~\ref{fig:SplittingProbabilities}b in the range $x_0 \gtrapprox -L/4$). By summing suitable combinations of the conditional splitting probabilities, i.e.\ $\pi(x_0,y_0) = \pi_{+}(x_0,y_0) + \pi_{-}(x_0,y_0)$, where $\pi(x_0,y_0)$ is the probability to observe \emph{any} particle state at the left-hand boundary given initialisation at $x_{0}$ in the $y_0 = \pm 1$ state, we can verify our results against the unconditional splitting probabilities derived in Ref.~\cite{malakar2018steady}, see Fig.~\ref{fig:SplittingProbabilities}c.

As discussed in Sec.~\ref{s:formalism}, these results may be combined with Bayes' theorem to determine the conditional splitting probabilities, i.e.\ the posterior likelihood of a particle's internal state, $y_{\rm exit} = \pm 1$, \emph{given} it has exited the interval $\mathbb{L}$ through a particular boundary. For instance, the probability of an RnT particle being a left mover given it was initialised at $x_0=0$ and has crossed the left-hand boundary at $x=-L/2$ is 
\begin{equation}\label{app:eq:BayesianSplitUpdate}
    \begin{split}
        \Pi(y_{\rm exit} = -1 | \text{left exit}, x = 0 )
        &= \frac{P(y_{\rm exit} = -1 \cap \text{left exit} )}{ P(\text{left exit} ) }\\
        &= \frac{ \pi_{-}(0,-)P_{\rm init}(y_0 = -1) + \pi_{-}(0,+)P_{\rm init}(y_0 = +1) }{ \pi(0,-) P_{\rm init}(y_0 = -1) + \pi(0,+) P_{\rm init}(y_0 = +1) }~,
     \end{split}
\end{equation}
where $P_{\rm init}(y_0 = \pm1)$ denotes the prior probability of the self-propulsion mode at initialisation. Similarly,
\begin{subequations}\label{eq:rnt_split_cond_all}
\begin{small}
\begin{align}
    \Pi(y_{\rm exit} = -1 | \text{right exit}, x = 0 )
        &= \frac{ \Pi(L/2, -|0,-)P_{\rm init}(y_0 = -1) + \Pi(L/2, -|0,+)P_{\rm init}(y_0 = +1) }{ (1-\pi(0,-)) P_{\rm init}(y_0 = -1) + (1-\pi(0,+)) P_{\rm init}(y_0 = +1) }~,  \\
    \Pi(y_{\rm exit} = +1 | \text{left exit}, x = 0 )
        &= \frac{ \pi_{+}(0,-)P_{\rm init}(y_0 = -1) + \pi_{+}(0,+)P_{\rm init}(y_0 = +1) }{ \pi(0,-) P_{\rm init}(y_0 = -1) + \pi(0,+) P_{\rm init}(y_0 = +1) }~,  \\
    \Pi(y_{\rm exit} = +1 | \text{right exit}, x = 0 )
        &= \frac{ \Pi(L/2,+|0,-)P_{\rm init}(y_0 = -1) + \Pi(L/2,+|0,+)P_{\rm init}(y_0 = +1) }{ (1-\pi(0,-)) P_{\rm init}(y_0 = -1) + (1-\pi(0,+)) P_{\rm init}(y_0 = +1) } ~.  \label{eq:rnt_split_cond}
\end{align}
\end{small}
\end{subequations}
In Ref.~\cite{cocconi2023optimal}, it was shown that a feedback control protocol based on this type of inference performs closely to the theoretical optimum in extracting work from an active system, despite the fact that the latter requires continuous tracking. 

\begin{figure}
    \centering
    \includegraphics[width=\textwidth]{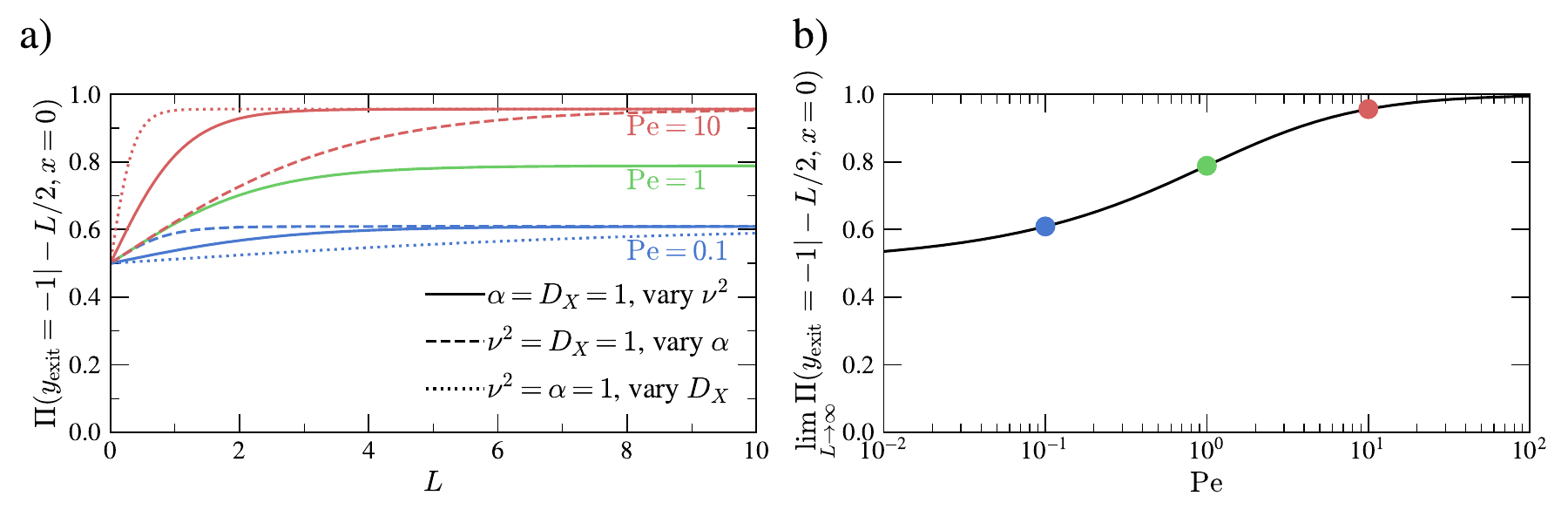}
    \caption{Conditional splitting probabilities for an RnT particle. a):
    Probability that an RnT particle initialised at $x_0 = 0$ exits the splitting window in the left-moving state \textit{given} that it exits through the boundary at $-L/2$, shown as a function of $L$ with the prior $P_{\rm init}(y_0 =+1)= P_{\rm init}(y_0 = -1) = 1/2$. Curves of the same colour correspond to RnT particles with the same P\'{e}clet number ${\rm Pe}= \nu^2/\alpha D \in \{0.1,1,10\}$. For each colour, the dashed, dotted and solid lines indicate different combinations of parameter values for $\nu$, $\alpha$ and $D_X$ that result in the same $\rm{Pe}$ number, e.g.\ solid lines indicate constant $\alpha = D_X = 1$ such that $\rm{Pe}$ is fully determined by $\nu^2$. For $\rm{Pe}= 1$, the dashed, dotted and solid lines coincide. 
   b): Plot of the asymptotic value ($L \to \infty$) of the conditional splitting probability as a function of $\rm{Pe}$. Large $\rm{Pe}$ corresponds to more persistent motion.
    }
    \label{fig:RnTinference}
\end{figure}

An example of the conditional splitting probabilities from Eq.~\eqref{eq:rnt_split_cond_all} is shown in Fig.~\ref{fig:RnTinference} as a function of the interval size $L$ and for different values of the P\'{e}clet number $\rm{Pe}=\nu^2/\alpha D$. The stronger persistence associated with higher $\rm{Pe}$ generically leads to a more noticeable deviation of the posterior probability from the symmetric prior $P(y_{\rm exit})=1/2$, indicating that the internal state can be inferred more reliably. Indeed, increasing $\rm Pe$ by decreasing the diffusivity $D$ leads, in the limit $D \to 0$, to dynamics where the left (respectively, right) boundary may be crossed only by left- (respectively, right-) moving particles.
For $L \to 0$, the dynamics are dominated by symmetric diffusion and no inference is possible. On the other hand, we find that the conditional splitting probability grows monotonically with $L$ to an asymptotic value that is controlled only by $\rm Pe$ (Fig.~\ref{fig:RnTinference}a). The large $L$ asymptote approaches 1 as ${\rm Pe} \to \infty$, allowing for exact inference, and reduces to $1/2$ as ${\rm Pe} \to 0$. This is a remarkable finding, indicating that while activity may amount to a mere renormalisation of the diffusivity in the bulk, $D_{\rm eff} = D + \nu^2/\alpha$, the persistence of RnT motion still plays an important role when observables are associated with boundary interactions.

\subsection{Intermittent piecewise-linear potential}\label{ss:intermittent}

As a second example, we consider one-dimensional Brownian motion in an intermittent potential $V(X)$, 
which switches ``on'' and ``off'' stochastically in the manner of a telegraph process with symmetric Poisson rate $r$. The dynamics are governed by the Langevin equation $\dot{X}(t) = -\mu Y(t)V'(X(t)) + \sqrt{2D}\xi(t)$, with $Y \in \{0,1\}$ constituting the hidden binary internal state that governs whether the potential is present or not at time $t$. 
Following the previous section, we introduce the compact notation for the joint splitting probability for exit at the left boundary $\pi_{y_{\rm exit}}(x_0,y_0) \equiv \Pi(-L/2,y_{\rm exit}|x_0,y_0)$ with $y_{\rm exit}, y_0 \in \{{\rm on} \equiv 1,{\rm off} \equiv 0\}$. These probabilities satisfy the coupled Kolmogorov backward equations 
\begin{align} \label{eq:FluctuatingPot}
        0 &= \Dx \frac{\partial^2 \pipm(x,1)}{\partial x^2} - V'(x) \frac{\partial \pipm(x,1)}{\partial x} + r[\pipm(x,0) - \pipm(x,1)]~, \nonumber \\
        0 &= \Dx \frac{\partial^2 \pipm(x,0)}{\partial x^2} + r[\pipm(x,1) - \pipm(x,0)]~,
\end{align}
where we have defined $y_1 \equiv y_{\rm exit}$ for notational ease and set the mobility $\mu$ to unity by suitable choice of units. The probabilities also satisfy the boundary conditions:
\begin{equation}
    \pipm(-L/2, y_0) = \delta_{y_1, y_0}, \hspace{0.3cm} \pipm(L/2, y_0) = 0~.
\end{equation}
In the following, we specialise to a piecewise-linear (ratchet) potential  with maximum barrier height $h \in \mathbb{R}$ and apex location $a \in [-L/2,L/2]$ of the form \cite{roberts2023run, roberts2024ratchet}
\begin{equation}\label{eq:piecewise_pot}
    V(x) = 
    \begin{cases}
        \frac{h}{\frac{L}{2} + a}\left(x+ \frac{L}{2}\right), \quad -\frac{L}{2} \leq x \leq a\\
        -\frac{h}{\frac{L}{2} - a}\left(x-\frac{L}{2}\right), \quad a \leq x \leq \frac{L}{2}
    \end{cases}~.
\end{equation}
With this definition of the potential, the problem is symmetric under the inversion $x \rightarrow -x$ and $a \rightarrow -a$. Thus, $\Pi( -L/2, y_1| x_0, y_0;a) = \Pi( L/2, y_1| -x_0, y_0; -a)$, motivating our choice to solve explicitly for $\pi_{y_1}(x_0,y_0)$ only. 
Given the piecewise structure of the potential, it is natural to also split $\pipm(x_0, 1)$ and $\pipm(x_0,0)$ into two regions:
\begin{equation}
    \pipm(x,1) = 
     \begin{cases}
        \pipmone(x,1), \quad -\frac{L}{2} \leq x \leq a\\
        \pipmtwo(x,1), \quad a \leq x \leq \frac{L}{2}
    \end{cases}~.
\end{equation}
The equations for all the degrees of freedom are given by 
\begin{subequations}
\begin{gather}
        \Dx \frac{d^2 \pipmone(x,1)}{d x^2} - \alpha \frac{d \pipmone(x,1)}{d x} + r[\pipmone(x,0) - \pipmone(x,1)] = 0~, \\
        \Dx \frac{d^2 \pipmone(x,0)}{d x^2} + r[\pipmone(x,1) - \pipmone(x,0)] = 0~, \\
        \Dx \frac{d^2 \pipmtwo(x,1)}{d x^2} + \beta\frac{d \pipmtwo(x,1)}{d x} + r[\pipmtwo(x,0) - \pipmtwo(x,1)] = 0~, \\
        \Dx \frac{d^2 \pipmtwo(x,0)}{d x^2} + r[\pipmtwo(x,1) - \pipmtwo(x,0)] = 0~,
    \end{gather} \label{eq:couped_pm_ratchet}
\end{subequations}
where we defined $\alpha = h/(L/2 + a)$ and $\beta = h/(L/2 - a)$. The boundary conditions now read
\begin{equation}\label{eq:bcforpi_also}
    \pipmone(-L/2, y_0) = \delta_{y_1 y_0}, \hspace{0.3cm} \pipmtwo(L/2, y_0) = 0, \hspace{0.3cm}
\end{equation}
and
\begin{equation}
\label{eq:bcforpi}
        \pipmone(a,y_0) = \pipmtwo(a,y_0), \hspace{0.5cm} \left. \frac{d \pipmtwo(x,y_0)}{dx} \right|_{x = a} = \left. \frac{d \pipmone(x,y_0)}{dx} \right|_{x = a} ~.
\end{equation}

To solve these equations, we first consider the equations relevant for the interval $x \in[-L/2,a]$. Once solved, it is clear by inspection of Eq.~\eqref{eq:couped_pm_ratchet} that one can replace $\alpha$ by $-\beta$ to recover the general solution in $x \in [a,L/2]$.  We define two useful quantities $\rhopmone(x) = \pipmone(x,1) + \pipmone(x,0)$ and $\sigmapmone(x) = \pipmone(x,1) - \pipmone(x,0)$. In terms of these quantities, the differential equations \eqref{eq:couped_pm_ratchet} become
\begin{subequations}
\label{eq:rhoandsigma}
    \begin{gather}
    \label{eq:rhoandsigmaa}
        \frac{d^2 \rhopmone(x)}{d x^2} - \frac{\alpha}{2 D} \frac{d \rhopmone(x)}{d x} - \frac{\alpha}{2 D}\frac{d \sigmapmone(x)}{d x} = 0~, \\
        \label{eq:rhoandsigmab}
        \frac{d^2 \sigmapmone(x)}{d x^2} - \frac{\alpha}{2 D} \frac{d \rhopmone(x)}{d x} - \frac{\alpha}{2 D}\frac{d \sigmapmone(x)}{d x} - \frac{2r}{D} \sigmapmone(x) = 0 ~ .
    \end{gather}
\end{subequations}
Using Eq.~\eqref{eq:rhoandsigmaa}, Eq.~(\ref{eq:rhoandsigmab}) can be written as 
\begin{equation}
\label{eq:sigmaderiv}
     \frac{d^2 \sigmapmone(x)}{d x^2} - \frac{d^2 \rhopmone(x)}{d x^2} - \frac{2r}{D} \sigmapmone(x) = 0~.
\end{equation}
Differentiating Eq.~(\ref{eq:rhoandsigmaa}) then gives an expression for the second derivative of $\sigmapmone(x)$:
\begin{equation}
    \frac{\alpha}{2 D} \frac{d^2 \sigmapmone}{dx^2} = \frac{d^3 \rhopmone}{dx^3} - \frac{\alpha}{2 D}\frac{d^2 \rhopmone}{dx^2}.
\end{equation}
Substituting this into Eq.~(\ref{eq:sigmaderiv}) then gives
\begin{equation}\label{eq:prefixingsigma}
    \frac{d^3 \rhopmone}{dx^3} - \frac{\alpha}{ D}\frac{d^2 \rhopmone}{dx^2} - \frac{\alpha r}{D^2} \sigmapmone(x) = 0~.
\end{equation}
Finally, integrating Eq.~\eqref{eq:rhoandsigmaa} allows us to write $\sigmapmone$ in terms of $\rhopmone$ and its derivatives:
\begin{equation}
\label{eq:fixingsigma}
    \frac{d \rhopmone(x)}{dx} = \frac{\alpha}{2 D} \left( \rhopmone(x) + \sigmapmone(x) - \czero \right)~.
\end{equation}
Combining  Eqs.\,\eqref{eq:prefixingsigma} and \eqref{eq:fixingsigma}, we find
\begin{equation}\label{eq:solS}
    \frac{d^3 \rhopmone}{dx^3} - \frac{\alpha}{ D}\frac{d^2 \rhopmone}{dx^2} - \frac{2 r}{D^2} \frac{d \rhopmone(x)}{dx} + \frac{\alpha r }{D^2} \rhopmone(x) = \frac{\alpha r }{D^2} \czero ~.
\end{equation}
Making the ansatz $\rhopmone(x) = e^{\lambda x} + ax^3/3! + bx^2/2  + cx +d$, where the exponential part solves for the complementary function, we find that $a=b=c=0$ and $d = \czero$, whereby 
\begin{equation}\label{eq:ratchet_rho_1}
    \rhopmone(x) = \czero + \cone e^{k_1 x} + \ctwo e^{k_2 x} + \cthree e^{k_3 x}~,
\end{equation}
where $k_i$ satisfies the characteristic equation
\begin{equation}
    k_i^3 - \frac{\alpha}{D}k_i^2 - \frac{2r}{D} k_i + \frac{\alpha r}{D^2} = 0 ~.
\end{equation}
Substituting the result for $\rho_s^{(1)}(x)$ into Eq.~(\ref{eq:fixingsigma}), we obtain the following solution for $\sigma_s^{(1)}(x)$ for $x\in[-L/2, a]$: 
\begin{equation}\label{eq:ratchet_sigma_1}
    \sigmapmone(x) = \cone \left(\frac{2 D k_1}{\alpha} - 1 \right) e^{k_1 x} + \ctwo \left(\frac{2 D k_2}{\alpha} - 1 \right) e^{k_2 x} + \cthree \left(\frac{2 D k_3}{\alpha} - 1 \right) e^{k_3 x} ~.
\end{equation}
To obtain the solution for the remaining interval $x \in [a,L/2]$, we can simply set $\alpha = -\beta$, resulting in
\begin{subequations}
\label{eq:solutionsigmarho}
    \begin{gather}
    \rhopmtwo(x) = \dzero + \done e^{k_1 x} + \dtwo e^{k_2 x} + \dthree e^{k_3 x}~, \\
    \sigmapmtwo(x) = -\done \left(\frac{2 D \tilde{k}_1}{\beta} + 1 \right) e^{ \tilde{k}_1 x} - \dtwo \left(\frac{2 D \tilde{k}_2}{\beta} + 1 \right) e^{\tilde{k}_2 x} - \dthree \left(\frac{2 D \tilde{k}_3}{\alpha} + 1 \right) e^{\tilde{k}_3 x} ~,
    \end{gather}
\end{subequations}
where $\tilde{k}_i$ satisfies
\begin{equation}\label{eq:solE}
     \tilde{k}_i^3 + \frac{\beta}{D}\tilde{k}_i^2 - \frac{2r}{D} \tilde{k}_i - \frac{\beta r}{D^2} = 0 ~.
\end{equation}
Re-expressing the solutions, Eqs.~\eqref{eq:ratchet_rho_1}, \eqref{eq:ratchet_sigma_1} and \eqref{eq:solutionsigmarho}, in terms of $\pipm^{(1,2)}$, we finally obtain
\begin{small}
\begin{subequations}
    \label{eq:solutionpi}
    \begin{alignat}{4}
    \pipmone(x,1) &= \frac{\czero}{2} + \frac{D k_1 \cone}{\alpha} e^{k_1 x} + \frac{D k_2 \ctwo}{\alpha} e^{k_2 x} + \frac{D k_3 \cthree}{\alpha} e^{k_3 x}~, \\
    \pipmone(x,0) &=  \frac{\czero}{2} + \cone \left(1 - \frac{D k_1}{\alpha} \right) e^{k_1 x} + \ctwo \left( 1 - \frac{D k_2}{\alpha}  \right) e^{k_2 x} + \cthree \left(1 - \frac{D k_3}{\alpha} \right) e^{k_3 x}~, \\
    \pipmtwo(x,1) &= \frac{\dzero}{2} - \frac{D \tilde{k}_1 \done}{\beta} e^{\tilde{k}_1 x} - \frac{D \tilde{k}_2 \dtwo}{\beta} e^{\tilde{k}_2 x} - \frac{D \tilde{k}_3 \dthree}{\beta} e^{\tilde{k}_3 x}~, \\
    \pipmtwo(x,0) &= \frac{\dzero}{2} + \done \left(\frac{D \tilde{k}_1}{\beta} + 1 \right) e^{ \tilde{k}_1 x} + \dtwo \left(\frac{ D \tilde{k}_2}{\beta} + 1 \right) e^{\tilde{k}_2 x} + \dthree \left(\frac{D \tilde{k}_3}{\alpha} + 1 \right) e^{\tilde{k}_3 x}~.
    \end{alignat}
\end{subequations}
\end{small}

\begin{figure}
    \centering
    \includegraphics[width=\linewidth]{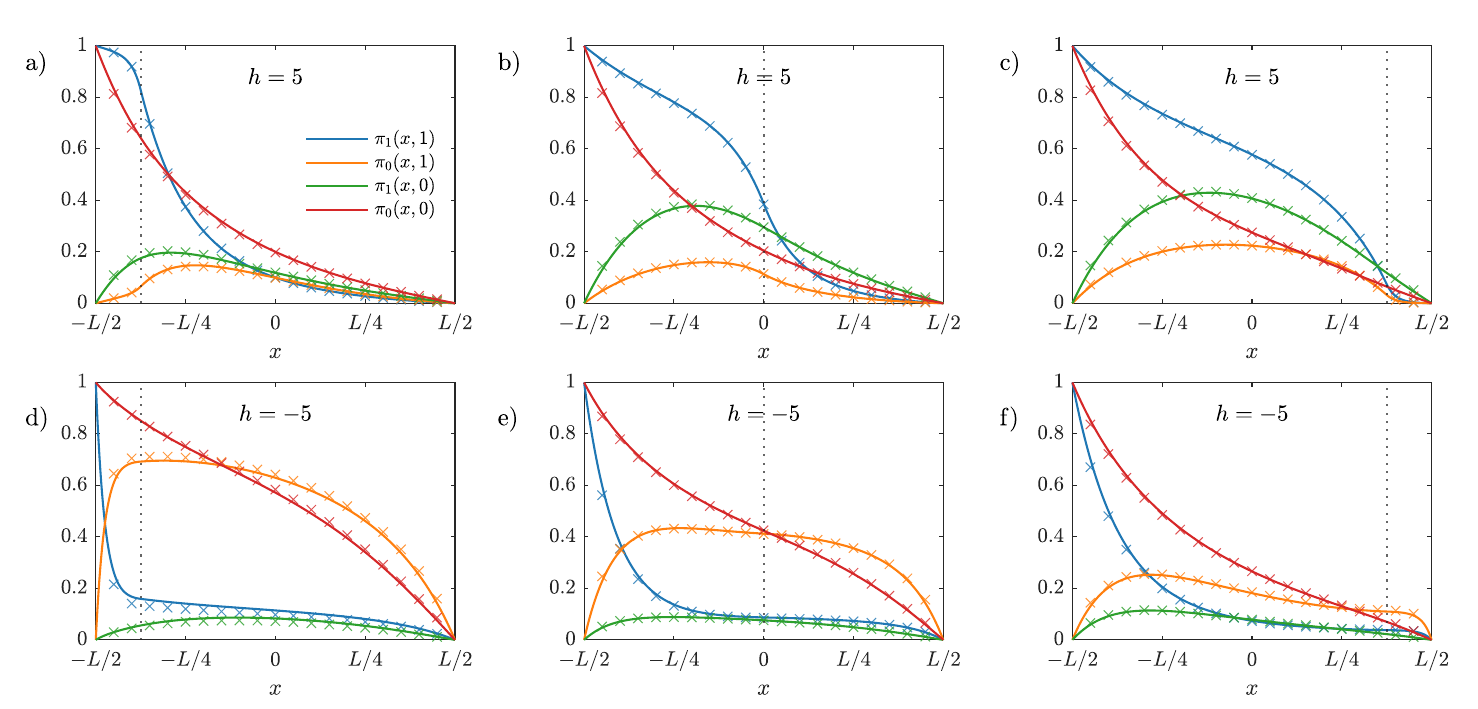}
    \caption{ Joint splitting probabilities, as a function of initialisation position $x$, for a Brownian particle subject to the intermittent piecewise-linear external potential \eqref{eq:piecewise_pot} crossing the left-hand boundary at $x_{\rm min}=-L/2$ where $L=4$ and $r=1$. Markers represent simulation results. Vertical dotted lines indicate the apex location $a$.  
    }
    \label{fig:joint_split_ratchet}
\end{figure}

All that remains to fix the integration coefficients is to apply the boundary conditions in Eqs.~\eqref{eq:bcforpi_also} and \eqref{eq:bcforpi}. Since the resulting expressions are neither compact nor particularly informative, we instead compute these coefficient numerically and show the resulting splitting probabilities graphically in Fig.~\ref{fig:joint_split_ratchet}, where we plot them as a function of $x$ for different values of the barrier height $h$ and apex location $a$.  
For $h > 0$, a particle on either side of the apex experiences a force pushing it outwards towards the boundaries during the ``on'' phase. For relatively small rates $r$ (infrequent switching), the probability $\pi_1(x,1)$ (plotted in blue) of exiting from the left boundary while the potential is switched on is thus enhanced for $x<a$  and reduced when $x>a$, compared to the case captured by $\pi_1(x,0)$ (plotted in green) where the potential is initially ``off'' (Fig.~\ref{fig:joint_split_ratchet}a--c). 
For $h < 0$ (resulting in a ``trapping" potential), activation of the potential hinders the escape of the particle from the interval and most exit events occur during the ``off'' phase (Fig.~\ref{fig:joint_split_ratchet}d--f, red and orange curves). Moreover, shifting the potential minimum $a$ towards one of the boundaries, reduces the overall probability of escaping from the other boundary. We argue that this is a consequence of the trapping potential biasing the relaxation of the distribution toward its minimum (Fig.~\ref{fig:joint_split_ratchet}d--f), while the change in the steepness with $a$ appears to play a minor role.

\begin{figure}
    \centering
    \includegraphics[width=0.9\linewidth]{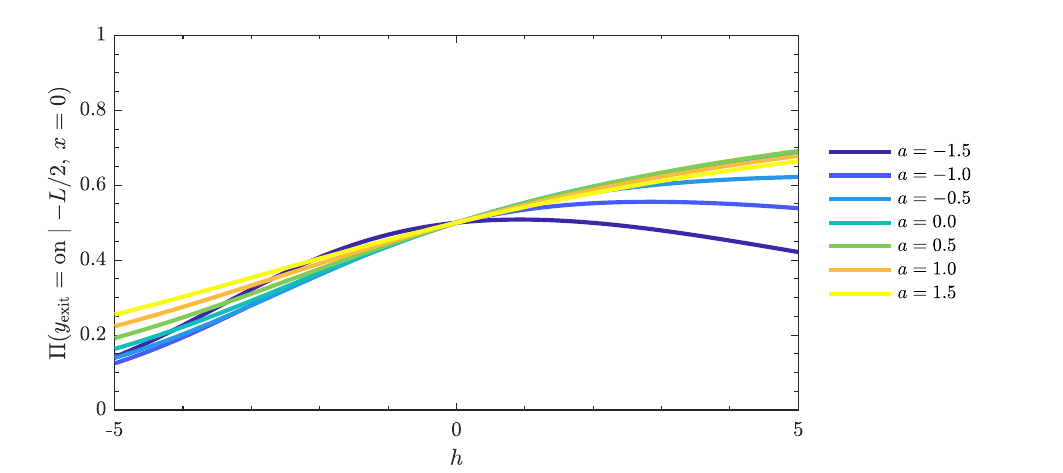}
    \caption{Conditional splitting probabilities for Brownian motion in the intermittent piecewise-linear external potential \eqref{eq:piecewise_pot} with the prior $P_{\rm init}(y_0 =0)= P_{\rm init}(y_0 = 1) = 1/2$, here shown as a function of the barrier height $h$ and for different values of the apex location $a$. The other parameter values are the same as those used in Fig.~\ref{fig:joint_split_ratchet}.
    }
    \label{fig:intermittentinference}
\end{figure}

The conditional form of these splitting probabilities can once again be obtained by drawing on Bayes' theorem, see Sec.~\ref{s:formalism}. An example of these is plotted in Fig.~\ref{fig:intermittentinference} for different values of the barrier height $h$ and apex location $a$, while $x=0$ is kept constant. As expected, the conditional splitting probability reduces to the symmetric prior $P(y_{\rm exit}) = 1/2$ for vanishing barrier height, $h \to 0$. 
For $h<0$, the potential is confining and the conditional probability of the potential being ``on'' given that a particle has just exited the interval from the left boundary is always less than $1/2$ for a particle initialised at $x=0$. This reflects the fact that the escape is generally hindered by the potential. Keeping $h$ constant, this probability changes non-monotonically with $a$: as $a \to -L/2$, diffusion up the confining potential to the left boundary becomes more likely as trapped particles will spend more time in the proximity of the latter. 
In the other limit, namely $a\to L/2$ we also see an increase in the conditional splitting probability. We argue that this is due to a reduction in the probability of escaping to the left boundary in the  ``off" state: the red and orange curves at $x=0$ reduces more sharply than the blue and green curves in Fig.~\ref{fig:joint_split_ratchet}d--f as $a$ increases.   
For $h>0$, we also see a (weakly) non-monotonic dependence on $a$. 
This can be rationalised as a competition between two effects: on one hand, setting $a>x$ ensures that potential activation drives the particle towards the left boundary; on the other, further increasing $a$ reduces the potential gradient and thus its effectiveness in biasing diffusion. This can be seen also in Fig.~\ref{fig:joint_split_ratchet}a--c where, as $a$ increases, the $x=0$ points on the blue and green curves increase, signifying increased likelihood of escape in the ``on" state. In Fig.~\ref{fig:joint_split_ratchet}c however, the $x=0$ point on the red and orange curves is also increased, suggesting that the potential pushes the particle towards the left boundary, but the potential may switch to the ``off" state before the particle escapes at $-L/2$. 

\subsection{Stochastic Resetting}\label{ss:resetting}
In the previous section, we focused on a particle subject to an intermittent, piecewise-linear potential. While this specific form was chosen for analytical convenience, the governing equations, Eqs.~\eqref{eq:FluctuatingPot}, remain valid for arbitrary potentials. However, solving the problem for a general potential quickly becomes intractable. To make further progress, we may consider a simplification that allows us to treat a wider class of systems---specifically, convex confining potentials with a single well-defined minimum in the interval $\mathbb{L}$.
If such potentials are also very stiff, the particle will be tightly localized around the minimum whenever the potential is active. In this limiting case and assuming that the deactivation rate is large, the activation of the potential $V(x)$ effectively corresponds to an instantaneous reset of the particle to a fixed position $x_r={\rm argmin}_\mathbb{L}[V(x)]$ \cite{alston2022non}.

With this motivation in mind, we now move to our final example of drift-diffusive systems with unidirectional coupling to a hidden state, namely one-dimensional Brownian motion subject to stochastic resetting \cite{evans2011diffusion,evans2020stochastic,gupta2020stochastic} with constant rate $r$. 
To keep our treatment as general as possible, we allow the post-reset position $x_r \in \mathbb{L}$ to be drawn from a distribution $P_r(x_r)$. In this case, we consider the hidden degree of freedom $Y(t)$ to be a counting process tracking the number of resetting events that the particle has undergone after its initialisation, whence $y_0 =0$.
The observable process $X(t)$ obeys the Langevin equation
\begin{equation}
    X(t +  dt) = 
    \begin{cases}
        X(t) +\sqrt{2D}\int_{t}^{t + dt}dt'~ \xi(t'), \hspace{0.1cm} \text{with probability} \hspace{0.2cm} 1-rdt \\
        x_r, \hspace{0.1cm} \text{with probability} \hspace{0.2cm} rdt
    \end{cases}
\end{equation}
where $x_r$ is drawn from $P_r(x_r)$. 

We are interested in the joint probability of $X$ exiting the interval $\mathbb{L}$ through a given boundary after being initialised at $x_0$, having undergone exactly $n$ reset events since initialisation.
We denote this quantity with the compact notation $\pi^{\sigma}_n(x_0) \equiv \Pi(\sigma L/2, n|x_0, y_0=0)$ where $\sigma \in \{-1,1\}$. 
These probabilities satisfy the coupled backward Kolmogorov equations
\begin{subequations} \label{eq:BKEresetting3}
\begin{gather}
     D\frac{d^2 \pisigma_0(x) }{d x^2} - r \pisigma_0(x)  = 0~,  \\
     D\frac{d^2 \pisigma_n(x) }{d x^2} - r \pisigma_n(x) + r \int_{-L/2}^{L/2} dx'~P_r(x')\pisigma_{n-1}(x') = 0, \hspace{0.3cm} n\geq 1~,
\end{gather}
\end{subequations}
with boundary conditions (in the fully permeable case)
\begin{equation}\label{eq:bc_resetting}
    \pisigma_n(\sigma'L/2) = \delta_{n,0} \delta_{\sigma \sigma'}~.
\end{equation}
To solve these equations, we first consider $\pisigma_0(x)$ since it is independent of $\pisigma_n(x)$ for $n \geq 1$. The general solution for this probability is given by 
\begin{equation}
    \pisigma_0(x) = A e^{-\alpha x} + B e^{\alpha x} ~,
\end{equation}
where we have defined the inverse length-scale induced by the resetting as $\alpha \equiv \sqrt{r/D}$. Applying the boundary conditions \eqref{eq:bc_resetting}, we find
\begin{equation}\label{eq:noreset_joint}
    \pisigma_0(x) = \frac{\sinh\left[\alpha \left(\frac{L}{2} + \sigma x \right)\right]}{\sinh(\alpha L)}~.
\end{equation}
Having solved the joint splitting probability in the case of no resetting, we can draw on it to calculate the joint splitting probability in the case of resetting.
Denoting $c^{\sigma}_{n-1} \equiv \int_{-L/2}^{L/2} dx'P_r(x') \pisigma_{n-1}(x')$, the general solution for the associated backward Kolmogorov equation is
\begin{equation}
    \pisigma_n(x) = A e^{\alpha x} + Be^{-\alpha x} + c^{\sigma}_{n-1}~.
\end{equation}
The uniformity of the boundary conditions, $\pisigma_n(L/2) = \pisigma_n(-L/2)=0$, implies $A = B$ and allows us to fix $A$, resulting in
\begin{equation} \label{eq:CSPresetting}
    \pisigma_n(x) = c^{\sigma}_{n-1} \left[ \frac{\cosh(\alpha L/2) - \cosh(\alpha x)}{\cosh(\alpha L/2)}\right]~.
\end{equation}
To find an explicit form for the prefactor $c^{\sigma}_{n-1}$, we substitute Eq.~\eqref{eq:CSPresetting} into its definition and perform the integral over the post-reset location to arrive at a recurrence relation 
between $c^{\sigma}_n$ and $c^{\sigma}_{n-1}$: \\
\begin{equation}
    c_n^{\sigma} = c^{\sigma}_{n-1} \left( 1 - \frac{\langle \cosh(\alpha x_r)\rangle}{\cosh(\alpha L/2)}\right) ~,
\end{equation}
where $\langle \cdot\rangle$ denotes an average over $P_r(x)$. Iterating this relation and using Eq.~\eqref{eq:noreset_joint}, we find
\begin{align}
    c^{\sigma}_n &= \left( 1 - \frac{\langle \cosh(\alpha x_r)\rangle}{\cosh(\alpha L/2)}\right)^n \int dx~P_r(x) \pisigma_n(x) \nonumber \\
    &= \frac{\langle \sinh(\alpha(L/2 + \sigma x_r))\rangle}{\sinh(\alpha L)}\left( 1 - \frac{\langle \cosh(\alpha x_r)\rangle}{\cosh(\alpha L/2)}\right)^n ~,
\end{align}
whence 
\begin{equation} \label{eq:fullCSPresetting3}
        \pisigma_{n}(x) = \frac{\langle \sinh\left(\alpha \left(\frac{L}{2} + \sigma x_r \right)\right) \rangle}{\sinh(\alpha L)} \left[1 - \frac{\langle \cosh(\alpha x_r) \rangle}{\cosh(\alpha L/2)}\right]^{n-1} \left[ 1 - \frac{\cosh(\alpha x)}{\cosh(\alpha L/2)}\right], \hspace{0.4cm}  n \geq 1 ~.
\end{equation}
Since $\pisigma_{n}(x)$ is a monotonically decreasing function of $n$, the most likely number of resets experienced by a particle that just exited the interval $\mathbb{L}$ is zero, independently of interval size or resetting rate. This may seem counter-intuitive, however it is consistent with resetting events being governed by a memoryless Poisson process, whereby each attempt to escape after a resetting event succeeds with probability $p_{\rm succ}<1$ independent of the number of previous failures. Importantly, the \emph{average} number of resetting events that occur before an exit event takes place does depend on both $\alpha$ and $L$, as expected.

\begin{figure}
    \centering
    \includegraphics[width=0.7\linewidth]{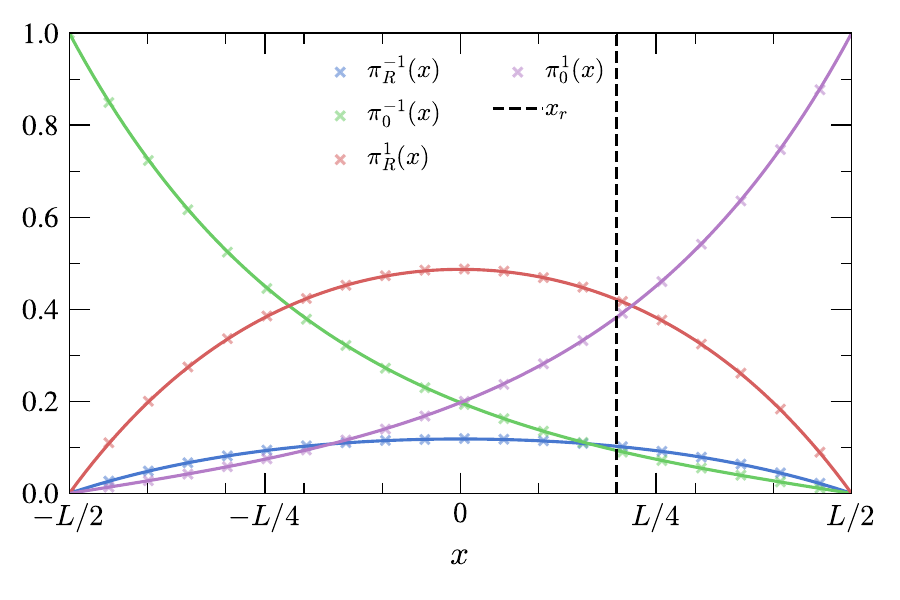}
    \caption{ Joint splitting probabilities $\pisigma_0(x)$, Eq.~\eqref{eq:noreset_joint}, and $\pisigmar(x)$, Eq.~\eqref{eq:CSPhavingreset}, for resetting Brownian motion initialised at $x$ to exit the interval $\mathbb{L}=[-L/2,L/2]$ through the boundary located at $x_{\rm exit}=\sigma L/2$ having undergone either no resetting events or at least one resetting event, respectively. The parameters values used in this figure are $D = 1, r = 2, L = 1$ and $x_r = 0.2$. The black dashed line indicates the resetting location while markers represent the results of simulations.}
    \label{fig:Resetting}
\end{figure}

Marginalising the joint splitting probabilities over $n\geq 1$ gives us the probability $\pisigma_R(x) \equiv \pisigma_{n\geq1}(x)$ of exiting from a given boundary while having reset at least once. 
The summation can be carried out exactly via a geometric sum, leaving us with the simple expression
\begin{equation} \label{eq:CSPhavingreset}
    \pisigmar(x) = \frac{\langle \sinh\left(\alpha \left(\frac{L}{2} + \sigma x_r \right)\right) \rangle}{\sinh(\alpha L)}  \left[ \frac{\cosh(\alpha L/2) - \cosh(\alpha x)}{\langle \cosh(\alpha x_r) \rangle} \right] ~.
\end{equation}
The joint splitting probabilities, Eqs.~\eqref{eq:noreset_joint} and \eqref{eq:CSPhavingreset}, are plotted in Fig.~\ref{fig:Resetting}. Equation~\eqref{eq:CSPhavingreset} suggests that, in order to maximise the probability of resetting at least once, the particle should be initialised at the midpoint of the interval, since this maximises the typical time before either boundary is reached for perfectly permeable boundaries. 
For a localised resetting distribution $P_r(x) = \delta(x-x_r)$, we also observe that the reset position $x_r$ that maximises the probability of leaving through boundary $x_{\rm exit}$ having reset at least once is $x_r=x_{\rm exit}$. 
Finally, we note that the joint probability $\pisigmar(x)$ is symmetric with respect to the initial condition $x$ and that it differs from $\pi_r^{-\sigma}(x)$ only by an $x$-independent factor.
This is because the initial position is relevant only insofar as it controls the probability of a resetting event occurring before a boundary is hit (which is itself symmetric), but is ``forgotten'' once the first resetting event takes place. Indeed, the probability of resetting before exiting may be computed explicitly to be
\begin{equation}
    p_{ \rm reset}(x) \equiv  \sum_{\sigma = \pm1} \pisigma_R(x)= \frac{2 \sinh(\alpha L/2)}{\sinh(\alpha L)} (\cosh(\alpha L/2) - \cosh(\alpha x)) ~,
\end{equation}
whence $\pisigmar(x) = \mathcal{G}(\alpha,L;P_r)p_{ \rm reset}(x)$.
 Said differently, after resetting, the relevant variable controlling the relative probability of exiting from either boundary is the resetting location $x_r$. This is most apparent for rare resets ($\alpha L \ll 1$), where the $x_r$-dependent term in Eq.~\eqref{eq:CSPhavingreset} reduces to the marginal splitting probability of Brownian motion initialised at $x_r$,
 \begin{equation}
     \pisigmar(x) \approx \left( \frac{1}{2} + \sigma \frac{\langle x_r \rangle}{L}\right)p_{\rm reset}(x) ~.
 \end{equation}

\begin{figure}
    \centering
    \includegraphics[width=0.9\linewidth]{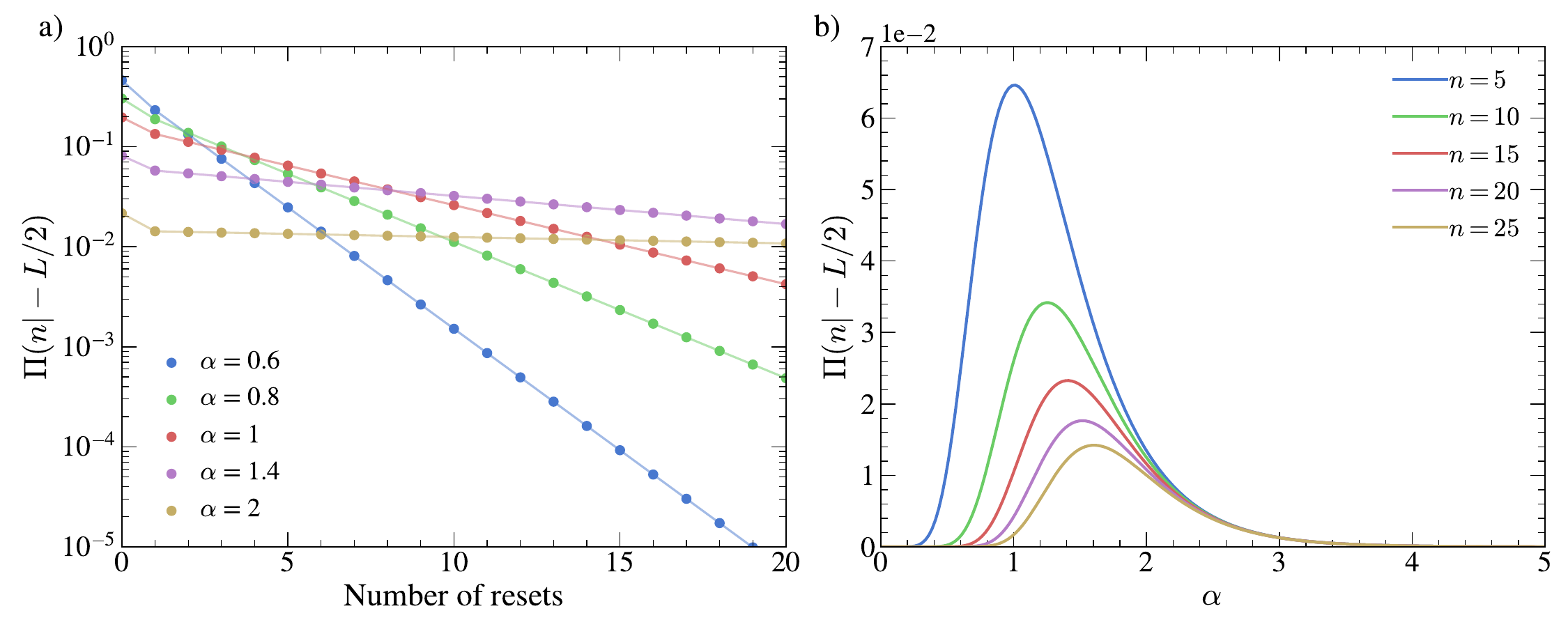}
    \caption{(a) Conditional splitting probability for a Brownian particle initialised at $x$ having undergone exactly $n \in \mathbb{N}$ resetting events before exiting the interval $\mathbb{L}$ from the left boundary $x_{\rm exit}=-L/2$, shown for different values of the characteristic inverse length scale $\alpha=\sqrt{r/D}$.
    The parameters values used in this figure are $L = 5$, $x_r = 0.2$ and $x_0 = 0$. (b) Conditional splitting probability as a function of $\alpha$ for different values of $n$.}
    \label{fig:resettinginference}
\end{figure}

Using Eq. (\ref{eq:cond_split_def2}), we can calculate the associated conditional splitting probabilities, namely the posterior likelihoods of the particle having undergone exactly $n$ resetting events before exiting, given that the exit occurred via a particular boundary. 
These are given by
\begin{equation} \label{eq:CSPresetting2}
    \Pi(n|\sigma L/2) = \frac{\pisigma_n(x)}{\sum_{n = 0}^{\infty} \pi^{\sigma}_n(x)} = \frac{\pisigma_n(x)}{\pisigma_0(x) + \pisigma_R(x)}
\end{equation}
and are plotted in Fig.~\ref{fig:resettinginference}, where we illustrate the asymptotic exponential decay with $n$ (cf.~Eq.~\eqref{eq:fullCSPresetting3}) with a characteristic rate controlled by $r$. Interestingly, the conditional splitting probability is a non-monotonic function of $\alpha$, implying the existence of an inverse length scale $\alpha^*(n)$ that maximises the probability of resetting exactly $n$ times before exiting.  
As before, while the most likely number of resets is zero independently of both $\sigma$ and $x_0$, the average number of resets conditioned on exit from a particular boundary depends non-trivially on these parameters. 
We can additionally define the conditional splitting probability of having reset at least once as
\begin{equation}
    \Pi(R|\sigma L/2) = \frac{\sum_{n = 1}^{\infty}\pisigma_n(x)}{\sum_{n = 0}^{\infty} \pi^{\sigma}_n(x)} = \frac{\pisigma_R(x)}{\pisigma_0(x) + \pisigma_R(x)}~.
\end{equation}

\section{Conclusion and Outlook} \label{s:conclusion}

Splitting probabilities constitute a standard tool in classical probability theory \cite{redner2001guide,van1992stochastic,hughes1996random,gardiner1985handbook}, with a number of applications ranging from population dynamics to chemical reactions and decision theory, as reviewed in Sec.~\ref{sec:intro}. For two-dimensional Markov processes $\{X(t),Y(t)\}_{t\in T}$, a joint analogue of the splitting probabilities can be defined, which captures the likelihood that the variables $X(t)$, having been initialised at $x_0 \in \mathbb{L}$, exits $\mathbb{L}$ for the first time via either of the interval boundaries \emph{and} that the internal state, initialised at $y_0$, is given by $y_{\rm exit}$ at the time of exit. 
Our first contribution in this work was to compute such joint splitting probabilities for a number of processes, which can be separated into two classes. 

The first class consists of processes where $X(t)$ is simple Brownian motion and $Y(t)$ is a decoupled (hidden) internal state, which may be discrete or continuous (Sec.~\ref{s:decoupled}). In this case, generic formulas for the joint probabilities can be derived, which are expressed in terms of the eigensystem of the Fokker-Planck operator associated with the $Y$ dynamics, Eqs.~\eqref{eq:joint_split_prob_semiperm_left} and \eqref{eq:joint_split_prob_semiperm_right}. For the sake of generality, we have derived these expressions assuming that either end-point of $\mathbb{L}$ may be a semi-permeable Robin boundary with permeability coefficient $\kappa \in [0,\infty)$. These general results are illustrated for two particular cases, namely one where $Y(t)$ is a discrete SIR-like process involving ripening and spoiling (Sec.~\ref{ss:rip_spoil}) and one where $Y(t)$ is a continuous Ornstein-Uhlenbeck process (Sec.~\ref{ss:ou}).

The second class consists of processes where $X(t)$ depends on $Y(t)$, while $Y(t)$ evolves independently (unidirectional coupling). For these processes, the joint splitting probabilities need to be computed on a case-by-case basis and we do so for three paradigmatic cases: free run-and-tumble motion (Sec.~\ref{ss:RnT_split}), Brownian motion in an intermittent piecewise-linear potential (Sec.~\ref{ss:intermittent}), and Brownian motion subject to stochastic resetting with an arbitrary resetting distribution (Sec.~\ref{ss:resetting}). 

Based on these results, our second contribution in this work was to introduce a recipe for inferring hidden internal states in multi-dimensional Markov processes by leveraging knowledge of their \emph{conditional} splitting probabilities. As defined in Sec.~\ref{s:formalism}, Eqs.~\eqref{eq:cond_split_def1} and \eqref{eq:cond_split_def2}, these are the posterior likelihoods of a hidden state \emph{given} that an observable degree of freedom has undergone a specific exit event. They can be derived straightforwardly from the joint splitting probabilities via Bayes' theorem and we have done so for all models studied in this work.
This approach enables the partial Bayesian inference of hidden variables from point-wise exit events \cite{cocconi2023optimal}.

When the dynamics of the hidden and observable degrees of freedom are decoupled, we show that inference based on exit events is theoretically possible only under non-stationary conditions (equivalently, for typical exit times smaller than the relaxation time of the hidden variable). On the other hand, inference remains possible at all times in cases where the observable degree of freedom depends on the hidden one. This difference is particularly striking in the limit of large intervals, $L\to \infty$, cf.\ the factorisation in Eq.~\eqref{eq:decoupled_factorisation} for the general decoupled problem and the nontrivial $L\to\infty$ asymptote of the posterior probability for RnT motion shown in Fig.~\ref{fig:RnTinference}a.

We suggest that this boundary-based inference scheme may find applications across fields concerned with the control of driven stochastic processes, thanks to its empirical implementation requiring only a mechanism for point-wise detection. For instance, in the context of active matter, the possibility of (partially) inferring the state of a motile particle is fundamental to the design of active information engines exploiting the persistence of self-propulsion forces \cite{cocconi2024efficiency,malgaretti2022szilard}. 

As already illustrated in Ref.~\cite{cocconi2023optimal}, the reliability of the inference may be improved by ``compounding'' evidence over multiple exit events, allowing for high accuracy even in regimes where the difference between the conditional splitting probability and the Bayesian prior for a single detection event is minimal. 

Finally, we note that, in all cases of unidirectional coupling between $X(t)$ and $Y(t)$ considered in this work, we have assumed that $Y(t)$ is a discrete variable. The calculation of joint and conditional splitting probabilities for continuous $Y(t)$ in this case, e.g.\ the one-dimensional active Ornstein-Uhlenbeck process \cite{bothe2021doi,martin2021statistical}, is an interesting open challenge which we leave for future work.  

\section*{Bibliography}
\bibliographystyle{iopart-num}
\bibliography{bibliography}

\end{document}